\documentclass[letterpaper,12pt]{article}

\usepackage{arxiv}

\usepackage[T1]{fontenc}    
\usepackage{hyperref}       
\usepackage{url}            
\usepackage{booktabs}       
\usepackage{amsfonts}       
\usepackage{nicefrac}       
\usepackage{microtype}      
\usepackage{lipsum}
\usepackage{graphicx}
\usepackage[T1]{fontenc} 
\usepackage{graphicx,color}
\usepackage[dvipsnames]{xcolor}
\usepackage{soul}
\usepackage{url}
\usepackage{subcaption}
\usepackage{amssymb}
\usepackage{float}
\usepackage{pdf14}
\usepackage{tikz}
\usepackage{placeins}

\usepackage{booktabs}  
\usepackage{amsmath}
\usepackage{textcomp}  

\usepackage{multirow}
\usepackage{rotating}
\usepackage{float}
\usepackage{rotating}
\usepackage{appendix}
\usepackage{comment}

\usepackage{amsthm}
\usepackage{multicol}

\usepackage{graphicx}
\usepackage{array}
\usepackage{multirow}
\usepackage{titlecaps, stringstrings}
\usepackage{float}
\usepackage{lscape} 

\usepackage{tabu}

\usepackage[utf8]{inputenc}

\usepackage{comment}
\usepackage{orcidlink}

\newcommand{\be}{\begin{equation}}
\newcommand{\ee}{\end{equation}}
\newcommand{\bea}{\begin{eqnarray}}
\newcommand{\eea}{\end{eqnarray}}

\usepackage{color}

\graphicspath{ {./fig/} }

\title{Educational Perspectives on Quaternions: Insights and Applications}

\author{
Fernando Ricardo González Díaz\orcidlink{0009-0009-9704-1119} \\
  CICATA-Legaria, Instituto Politécnico Nacional, Ciudad de México, 11500, México.\\ \texttt{fgonzalez33@outlook.com}
\And
  Vicent Martinez Badenes\orcidlink{0000-0002-4056-9627}\\
  Universidad Internacional de Valencia - VIU, 46002, Valencia, Spain.\\ \texttt{vicent.martinezb@professor.universidadviu.com}
\And 
  Ricardo García-Salcedo\orcidlink{0000-0003-0173-5466} \\
  CICATA-Legaria, Instituto Politécnico Nacional, Ciudad de México, 11500, México.\\
  and \\
  Universidad Internacional de Valencia - VIU, 46002, Valencia, Spain.\\
  \texttt{rigarcias@ipn.mx}
}

\begin{document}
\maketitle
\begin{abstract}
Quaternions, discovered by Sir William Rowan Hamilton in the 19th century, are a significant extension of complex numbers and a profound tool for understanding three-dimensional rotations. This work explores quaternion's history, algebraic structure, and educational implications. We begin with the historical context of quaternions, highlighting Hamilton's contributions and the development of quaternion theory. This sets the stage for a detailed examination of quaternion algebra, including their representations as complex numbers, matrices, and non-commutative nature. Our research presents some advancements compared to previous educational studies by thoroughly examining quaternion applications in rotations. We differentiate between left and right rotations through detailed numerical examples and propose a general approach to rotations via a theorem, clearly defining the associated morphism. This framework enhances the understanding of the algebraic structure of quaternions. A key innovation is the presentation of a three-dimensional example illustrating the rotation of a frame with strings, connecting quaternions to the quaternion group, half-integer spin phenomena, and Pauli matrices. This approach bridges theoretical concepts with practical applications, enriching the understanding of quaternions in scientific contexts. We emphasize the importance of incorporating the history and applications of quaternions into educational curricula to enhance student comprehension and interest. By integrating historical context and practical examples, we aim to make complex mathematical concepts more accessible and engaging for students at the undergraduate and graduate levels. Our study underscores the enduring relevance of quaternions in various scientific and technological fields and highlights the potential for future research and educational innovations.
\end{abstract}

\keywords{Quaternions \and Rotations \and  Algebraic structures \and Mathematical applications \and Educational research}


\section{Introduction}

Quaternions, developed by Sir William Rowan Hamilton in the 19th century, represent an extension of complex numbers to four dimensions and constitute a sophisticated framework for describing rotations and spatial orientations. Unlike real and complex numbers, which operate in one and two dimensions, respectively, quaternions exist in a four-dimensional space defined by their scalar and vector components. This peculiarity allows them to elegantly represent rotations in three-dimensional space without the ambiguities and singularities associated with other representations, such as Euler angles. Its distinctive properties include non-commutativity, which gives quaternion algebra a greater complexity than conventional algebraic systems. Despite this complexity, quaternions possess remarkable computational efficiency and numerical stability, making them indispensable tools in fields such as computer graphics \cite{vince2011quaternions, taubin20113d}, robotics, aerospace engineering \cite{wie1989quarternion, kristiansen2005satellite, filipe2015adaptive, valverde2018spacecraft} and physical simulations \cite{rapaport1985molecular, degond2018quaternions, zhu2022review, fu2022accurate}. By providing a solid and concise representation of spatial transformations, quaternions have become essential for engineers, scientists, and researchers from various disciplines who seek to model, analyze, and manipulate complex systems and phenomena with precision and accuracy.

This work aims, in part, to understand the importance of the history of mathematical concepts as a fundamental basis for understanding their development and evolution. It will explore the origin of quaternions in the work of William Rowan Hamilton \cite{lam2003hamilton}, appreciating the ingenuity behind their creation and how they emerged to address specific problems in geometry and algebra. Incorporating the history of quaternions into teaching can foster a richer and more complete understanding of these numbers, which may generate greater interest and learning in students. The target audience for this work comprises students in their final semesters of undergraduate and graduate programs.

A profound comprehension of quaternions, encompassing their algebraic structure and the mathematical operations defining them, is essential for unlocking their potential across diverse applications. 
This comprehension extends beyond mastering basic operations such as addition, subtraction, multiplication, and division. A more profound knowledge of concepts such as the conjugate, the norm, the multiplicative inverse, the dot product, and the cross product is required. By mastering these concepts, students can fully leverage the capabilities of quaternions in various scientific and technological fields \cite{mukundan2002quaternions, barry2016application, bayro2021survey}.

Finally, it is essential to provide simple examples of the application of quaternions for educational purposes. Presenting concrete cases in everyday situations and specific problems of computing and robotics allows students to become familiar with their utility and develop an intuition about their use in more advanced contexts. These examples not only consolidate the understanding of basic concepts but also prepare the ground for exploring more complex applications, such as those related to rotations in three-dimensional space.

This article aims to comprehensively understand quaternions, from their historical origin to their algebra and fundamental properties. To this end, practical and educational examples will illustrate their application in everyday contexts and specific areas. Additionally, it seeks to establish a solid foundation for future studies and applications of quaternions, providing students with the tools and knowledge necessary to tackle more advanced problems in fields such as physical simulation, computer animation, space navigation, and biomechanics.

\section{Exploring Quaternions in Teaching}

In recent years, the teaching of quaternions to university students has grown significantly, driven by their increasing relevance in various scientific and technological areas. However, despite their importance, the academic literature still lacks resources that offer a comprehensive understanding of quaternions at this educational level.

This section aims to fill this gap by offering a brief, non-exhaustive review of quaternion teaching, emphasizing effective pedagogical strategies and a systematic presentation of concepts. By exploring the literature related to quaternion teaching, this work emerges as a valuable resource for educators and students, strengthening the understanding and application of quaternions in both educational and professional realms.

From a methodological perspective, \cite{mcdonald2010teaching} proposed a constructive method for teaching quaternions, focused on developing intuition and understanding their application in rotation matrices. \cite{rodman2014topics} offered a comprehensive exposition of quaternion linear algebra, including applications across various fields. Additionally, \cite{da2021quaternios} presented educational software to facilitate quaternions' teaching and learning process, suggesting a potential path for enhancing understanding of this complex topic. Finally, \cite{markley2008unit} provides a modified method for extracting quaternions from rotation matrices.   

Regarding analogies for understanding quaternions, \cite{staley2010understanding} demystified the Dirac belt trick, a popular physics analogy for quaternions, highlighting the underlying four-dimensional parametric space for rotations. Similarly, \cite{diaz2017fenomeno} described additional examples of the Dirac belt phenomenon and related the algebraic structure of quaternions.

Regarding quaternion applications, \cite{henriksen2014relativity} discussed the challenges and opportunities in teaching quantum physics and relativity, closely related to understanding quaternions. On the other hand, \cite{bonacci2021teaching} and \cite{montgomery2022introduction} emphasize the importance of motivation and practical applications in teaching complex mathematical concepts like quaternions. Finally, \cite{kartiwa2023review} reviewed the historical development and applications of quaternion differential equations, providing a comprehensive overview of the subject.

This work aims to follow in the line of works such as those by \cite{familton2015quaternions} and \cite{furui2021understanding}, which provide a historical and theoretical context for the use of quaternions in physics, which can be valuable for university students to understand the importance of these concepts. 

Unlike the previous work of  \cite{mcdonald2010teaching}, our study comprehensively addresses the applications of rotation using quaternions. We present various representations, including complex numbers and matrices, and differentiate between left and right rotations through detailed numerical examples.

We propose a general approach to rotations through a theorem, clearly define the associated morphism, and discuss its physical significance. Innovatively, we present a three-dimensional example illustrating the rotation of a frame with strings, establishing connections between quaternions, the quaternion group, the phenomenon of half-integer spin, and the Pauli matrices.

This comprehensive and practical approach constitutes a significant contribution to the field of quaternions, their teaching at the higher education level, and their application in various scientific and technological fields.

\section{But, what are quaternions?} 

The geometric representation of a real number is a single point on an infinitely long straight line; this line has a defined unit: the distance between consecutive points representing the so-called integer numbers. The solution to the equation $x^2 + 1 = 0$ does not exist in the field of real numbers  $\mathbb{R}$, since the product of two positive real numbers is positive, and the product of two negative real numbers is also positive. However, by extending the set of real numbers, the set of complex numbers appears, constituting an extension of the real numbers.

\begin{figure}[h]
       \centering
       \includegraphics[width=7cm, height=7cm]{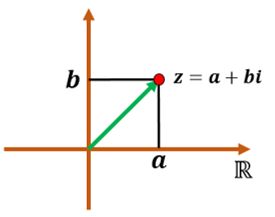}
        \caption{Geometric representation of a complex number $z$ in the Argand plane. Source: Author's elaboration.}
        \label{complejo}
\end{figure}

A complex number can be represented in different ways, some of which are as follows:

\begin{enumerate}
    \item Representation of a complex number as a point on a Cartesian plane  $z=(a,b)\in \mathbb{R}^2$. This plane has rectangular coordinates with two perpendicular axes, one horizontal and one vertical, called the real and imaginary axes, respectively. The real and imaginary parts are represented on their respective axes. The location of complex numbers is the same as the points in the Euclidean plane  $\mathbb{R}^2$.
    
    \item Representation of a complex number as a vector  $z=(a,b)$ directed line segment is located at the point in the Argand diagram (The term "Argand diagram" refers to the French mathematician and physicist Jean-Robert Argand, who introduced this type of graphical representation in the 19th century), as shown in Figure \ref{complejo}. The vector is formed from the origin to the previously located point.

    \item Representation of a complex number as the sum of two terms  $z=a+ib$, where $a$ and $b$ are real numbers ($a, b \in \mathbb{R}$), and $i$ is the imaginary unit, with the property $i^2 = -1$. The real part of $z$ is denoted as $Re(z) = a$, and the imaginary part as  $Im(z) = b$.
    
\end{enumerate}

The set of complex numbers is denoted as $\mathbb{C}$. The notion of a square root of  $-1$ allows the equation  $x^2 + 1 = 0$ to have a solution in complex numbers.

Complex numbers, a two-dimensional vector space over $\mathbb{R}$, are defined with the basis $B = \{1, i\}$. Arithmetic operations, such as addition, are defined component by component: $(a + bi) + (c + di) = (a + c) + (b + d)i$. The multiplication of two complex numbers  $z_1=a+bi$ and $z_2=c+di$ is carried out distributively using the property $i^2 = -1$:
\be
    z_1 z_2=(a+bi)(c+di)=(ac-bd)+(ad+bc)i.
\ee

Each complex number $z = a + bi$ has a conjugate, $z^\ast=\overline{z} = a - bi$, and it holds that $zz^\ast=z \overline{z} = |z|^2 = (a + bi)(a - bi) = a^2 + b^2$, representing the squared norm of $z$. 

Another way to represent a complex number is in its polar form. Recalling the equations for converting from rectangular to polar coordinates and adapting it to the Argand plane are $x=rcos\theta$ and $y=rsin\theta$, where $r$ is the distance from the origin to the point through a straight line (magnitude of the vector) and $\theta \in [0,2\pi]$ is the angle formed by the said line and the real axis. $\theta$ is the argument or phase and is denoted by $Arg(z)$ is $z$, the complex number it corresponds to. Substituting the equations of $x$ and $y$ in the definition of complex number we have, $$z=r(cos\theta + i sin\theta)$$

Recalling Euler's property, which establishes the equality  $e^{x+iy}=e^x(cosx+iseny)$ and substituting in the definition of a complex number, we have the following equalities:

\be
z=(a,b)=a+ib=r(cos\theta + i sin\theta)=re^{i \theta}
\ee

Additionally, any non-zero complex number has a multiplicative inverse  $z^{-1} = \frac{\overline{z}}{|z|^2}=\frac{z^\ast}{|z|^2}$. If the complex number  $z$ is unitary (has norm one), then it holds that the multiplicative inverse coincides with its conjugate $z^{-1}=z^\ast.$

These operations endow complex numbers with a structure of a normed division algebra  \cite{byrne2013}. A normed division algebra provides an algebraic structure in which the magnitude of the elements can be measured, and the existence of divisions for non-zero elements is guaranteed.

Quaternions $\mathbb{H}$ are an extension of the complex numbers $\mathbb{C}$. These numbers are called quaternions because they contain four basic vectors: $(1, i, j, k)$. According to the Merriam-Webster dictionary \footnote{https://www.merriam-webster.com/dictionary/quaternion}, it refers to a set of four parts (people or things). According to Hamilton's biographer, Robert Graves, Hamilton discussed the etymology of ''quaternion'' \cite{graves1889life}.

Using the analogy with the definition of complex numbers as the sum of two terms \( z = a + ib \) quaternions have a real part, \( \text{Re}(q) = a \), and an imaginary part, \( \text{Im}(q) = bi + cj + dk \), so that a quaternion \( q \) can be expressed as the sum of these two: \[ q = a + bi + cj + dk \] where \( a, b, c, d \in \mathbb{R} \) and the values \( i \), \( j \), and \( k \) are three distinct square roots of \( -1 \), which means they satisfy the algebraic property \( i^2 = j^2 = k^2 = ijk = -1 \).

\subsection{Basic Operations with Quaternions: Addition and Scalar Product }

Once we have defined quaternions, the next question arises: how are quaternions added? The addition of two quaternions is carried out by components, just like in complex numbers. Considering two quaternions $q_1 = a + bi + cj + dk$ and $q_2 = p + mi + rj + sk$, then the addition is defined as:
\be
    q_1+q_2=(a+bi+cj+dk)+(p+mi+rj+sk)=(a+p)+(b+m)i+(c+r)j+(d+s)k
\ee
The scalar product of a real number  $t \in \mathbb{R}$ by a quaternion $q_1 \in \mathbb{H}$ follows:
\be
    t\cdot q_1=t \cdot (a+bi+cj+dk)=ta+tbi+tcj+tdk
\ee
These operations satisfy associative, additive neutral, additive inverse, commutative, pseudoassociative, distributive, and scalar multiplicative neutral properties. Let the quaternions $q$, $q_1$, $q_2$ and $q_3$, then:
\begin{enumerate}
    \item Associative property of addition: $q_1 + (q_2 + q_3) = (q_1 + q_2) + q_3$
    \item Property of the additive neutral: $q + 0 = 0 + q = q$, where $0 = 0 + 0i + 0j + 0k$
    \item Property of the additive inverse: $q - q = -q + q = 0$, where the additive inverse is $-q = -a - bi - cj - dk$
    \item Commutative property of addition: $q_1 + q_2 = q_2 + q_1$
    \item Pseudoassociative property: $r(sq) = (rs)q$, where $r, s \in \mathbb{R}$
    \item Distributive properties: $r(q_1 + q_2) = (rq_1) + (rq_2)$ and $(r + s)q = rq + sq$
    \item Property of the scalar multiplicative neutral: $1q = q1 = q$
\end{enumerate}

With the abovementioned operations, quaternions $\mathbb{H}$ form a four-dimensional vector space over the field of real numbers $\mathbb{R}$. A four-dimensional vector space over the field of real numbers $\mathbb{R}$ is a set of elements (vectors) that satisfy certain axioms of linearity and operations with scalars. The dimension 4 indicates that any linearly independent set of four vectors in this space can generate the entire space. Since quaternions form a vector space and each quaternion is the sum of a real part and an imaginary part, the two concepts can be linked to define quaternions in the following way. The canonical orthonormal base for $\mathbb{R}^3$ is given by the three unit vectors $i=(1,0,0), j=(0,1,0), k=(0,0,1)$. A quaternion $q$ is written as the sum of a real scalar $a \in \mathbb{R}$ (using the analogy of the real part) and a vector  $\bold{q}=(b,c,d) \in \mathbb{R}^3$ (using the analogy of the imaginary part); in the form: 
\bea
    q&=&a+\bold{q} \nonumber \\
     &=&a+bi+cj+dk
\eea
\subsection{Quaternion Multiplication}
The multiplication of quaternions, which initially represented a challenge for Hamilton, is based on the distributive property and fundamental relationships between the imaginary units $i$, $j$, and $k$. These relationships are crucial for defining the product of quaternions and are expressed as:
\begin{equation}
    ij = k, \;\; jk = i, \;\; ki = j, \;\; ji = -k, \;\; kj = -i, \;\; ik = -j.
\end{equation}

Using these relationships, the product of two quaternions  $q_1 = a + bi + cj + dk$ and $q_2 = p + mi + rj + sk$ is defined as:
\begin{align}
    q_1 q_2 &= (a + bi + cj + dk)(p + mi + rj + sk) \nonumber \\
            &= (ap - bm - cr - ds) + (am + bp + cs - dr)i + (ar - bs + cp + dm)j + (as + br - cm + dp)k.
\end{align}

This process is simplified by using the dot product  $(\cdot)$ and the cross product  $(\times)$ of vectors in $\mathbb{R}^3$, allowing a more concise way to represent the product of quaternions:
\begin{equation}
q_1q_2 = ap - \mathbf{q}_1 \cdot \mathbf{q}_2 + a\mathbf{q}_2 + p\mathbf{q}_1 + \mathbf{q}_1 \times \mathbf{q}_2. \label{2daPC}
\end{equation}

\textbf{Example:} Consider $q_1 = 3 + i - 2j + k$ and $q_2 = 2 - i + 2j + 3k$.

\begin{enumerate}
    \item Quaternion product using the first definition:
    \bea
    q_1 q_2&=&(3+i-2j+k)(2-i+2j+3k) \nonumber \\
         &=&(3(2)-1(-1)-(-2)2-1(3))+(3(-1)+1(2)+(-2)3-1(2))i \nonumber \\ &+&(3(2)-1(3)+(-2)2+1(-1))j+(3(3)+1(2)-(-2)(-1)+1(2))k \nonumber \\
         &=&8-9i-2j+11k 
    \eea
    
    \item Quaternion product using the second definition: 
    
    The vector parts of the quaternions are $\bold{q_1}=(1,-2,1)$ and $\bold{q_2}=(-1,,2,3)$, respectively. Calculating the dot product and cross product, respectively, of the corresponding vectors, the following relation is obtained: 
    $$\bold{q_1}\cdot\bold{q_2}=(1,-2,1)(-1,2,3)=1(-1)+(-2)2+1(3)=-1-4+3=-2$$
    \\
    $$\bold{q_1}\times\bold{q_2}= 
    \left|\begin{smallmatrix}
        i & j & k\\
        1 & -2 & 1\\
        -1 & 2 & 3
\end{smallmatrix}\right|
       = [(-2)3-1(2)]i-[1(3)-1(-1)]j+[1(2)-(-2)(-1)]k=-8i-4j
    $$
    Using the formula of the second definition of quaternion product  (\ref{2daPC}), we have:
    
     \bea 
    q_1 q_2&=&6-(-2)+3(-i+2j+3k)+2(i-2j+k)+(-8i-4j) \nonumber \\
         &=&8-9i-2j+11k 
    \eea
\end{enumerate}

To describe the operation of multiplication in the set of quaternions $\mathbb{H}$, it is useful to remember some fundamental properties, including associative and distributive properties involving scalars  $r \in \mathbb{R}$ and quaternions  $q_1, q_2 \in \mathbb{H}$:

\begin{enumerate}
    \item \textbf{Associative property with scalars:} For every scalar $r\in \mathbb{R}$ and quaternions $q_1, q_2 \in \mathbb{H}$, it holds that
    \[
    (r q_1)q_2 = (q_1 r)q_2 = (q_1 q_2)r = q_1(r q_2).
    \]

    \item \textbf{Left distributive property:} 
    \[
    (q_1+q_2)q_3 = q_1 q_3 + q_2 q_3.
    \]

    \item \textbf{Right distributive property:} 
    \[
    q_1 (q_2+q_3) = q_1 q_2 + q_1 q_3.
    \]

    \item \textbf{Associative Property:} 
    \[
    q_1 (q_2 q_3) = (q_1 q_2) q_3.
    \]
\end{enumerate}

A crucial algebraic property of quaternion multiplication is that unlike multiplication in the sets of complex numbers  $\mathbb{C}$ or real numbers $\mathbb{R}$, it is not commutative. This non-commutativity has notable consequences, such as polynomial equations in  $\mathbb{H}$ may have more solutions than the degree of the polynomial. For example, the equation  $x^2 + 1 = 0$, which lacks a solution in  $\mathbb{R}$ and has two solutions in $\mathbb{C}$ ($\pm i$), has infinitely many solutions in $\mathbb{H}$. These include all points on the sphere's surface $S^3$ contained in $\mathbb{H}$.

To illustrate this statement, consider a quaternion  $q = a + bi + cj + dk$ whose square is $q^2 = -1$. In terms of $a, b, c,$ and $d$, , this translates into the system of equations:
\begin{equation*}
    a^2 - b^2 - c^2 - d^2 = -1, \quad 2ab = 0, \quad 2ac = 0, \quad 2ad = 0.
\end{equation*}

Solving this system, we find that  $a = 0$ and $b^2 + c^2 + d^2 = 1$. That is, the square of a quaternion is equal to $-1$ if and only if said quaternion is a pure imaginary of norm 1. The set of such quaternions forms the sphere $S^3$.

\subsection{The Conjugate, Norm, and Inverse of a Quaternion}

The operations of conjugation and division in quaternions are analogous to the corresponding operations in complex numbers. Consider a quaternion  $q = a + bi + cj + dk = a + \mathbf{q}$, where $\mathbf{q}$ represents the vector part of the quaternion. The conjugate of $q$ is defined as:
\begin{equation}
    q^{\ast} = a - bi - cj - dk = a - \mathbf{q} = \overline{q}.
\end{equation}

From the definition of the conjugate of a quaternion, the following properties can be well established:
\begin{enumerate}
    \item $(q^{\ast})^{\ast} = q$,
    \item $q + q^{\ast} = 2a$,
    \item $q^{\ast} q = qq^{\ast}$,
    \item $(q_1q_2)^{\ast} = q_2^{\ast} q_1^{\ast}$.
\end{enumerate}

The norm of a quaternion $q$ is defined as the scalar:
\begin{equation}
    \|q\| = \sqrt{q q^{\ast}}.
\end{equation}

A quaternion is called \textit{unitary} if its norm is equal to unity (1). Using the definition of norm, it holds that: 
\begin{equation}
    \|q_1q_2\|^2 = \|q_1\|^2\|q_2\|^2.
\end{equation}

Any non-zero quaternion has a multiplicative inverse, defined as: 
\begin{equation}
    q^{-1} = \frac{q^{\ast}}{\|q\|^2}.
\end{equation}

Using the definition of the multiplicative inverse, it can be verified that: 
\begin{equation}
    q^{-1}q = qq^{-1} = 1,
\end{equation}
and if the quaternion is unitary, then its inverse  $q^{-1}$ is equal to its conjugate  $q^{\ast}$.

\section{The Discovery of Quaternions by Hamilton and Their Historical Impact}

The discovery of quaternions is among the most meticulously recorded events in the history of mathematics  \cite{naiman1974role, buchmann2011brief, familton2015quaternions, vince2018quaternions}. Generally, having such detailed information about the date and place of a mathematical discovery of such importance is extremely unusual.

William Rowan Hamilton embarked on the quest for an extension of complex numbers  \cite{lam2003hamilton}, which proved fundamental for understanding rotations in two dimensions. His desire was to find a system that would address rotations in three dimensions, and although he had unsuccessfully attempted various approaches, it was during a walk from Dunsink to Dublin, Ireland, when Hamilton experienced a revelatory moment of intuition. He realized he could use a four-dimensional system to understand rotations in three dimensions, leading to what we know today as "quaternions." This concept has wide applications in fields such as satellite navigation and computer graphics generation  \cite{mukundan2002quaternions}.

In Hamilton's time, complex numbers represented by the mathematical set  $\mathbb{C}$ were a hot topic of research. An obvious question arose: if there was already a rule for multiplying two complex numbers, what happened with the multiplication of three numbers? \cite{leng2002phenomenology, lam2003hamilton}. This question haunted Hamilton for over a decade, and his son posed it at every breakfast. The answer to this question was linked to the non-existence of a three-dimensional normed division algebra  \cite{baez2002octonions}.

The solution to this problem arrived spectacularly on Friday, October 16, 1843. While Hamilton was walking with his wife Helen to the Royal Irish Academy, he suddenly conceived the idea of adding a fourth dimension to be able to multiply triples \cite{jia2008quaternions}. Excited by this breakthrough, when the couple passed the Brougham Bridge in Dublin\footnote{https://www.talesofawanderer.com/blog/2016/10/20/broom-bridge/}, he carved into a stone the fundamental equation of quaternions: $i^2=j^2=k^2=ijk=-1$. This equation symbolizes the culmination of Hamilton's efforts to extend complex numbers and understand rotations in three dimensions \cite{altmann1986rotations}. Although the original carving is no longer visible, a commemorative plaque at the site attests to this historic moment \cite{hamilton1854lxxvii}.

Driven by his conviction in quaternions, Hamilton founded a school dedicated to their study, called "Quaternionists"  \cite{ebbinghaus2012numbers}, and wrote an extensive treatise titled "Elements of Quaternions" \cite{hamilton1866elements}. However, by the end of the 19th century, quaternions were overshadowed by the vector analysis developed by Gibbs and Heaviside \cite{gibbs1893quaternions, heaviside1893vectors}. 

Although Hamilton did not see all the applications of quaternions in his time, he left a lasting mathematical legacy. His efforts to reconcile space and time through quaternions, influenced by the ideas of Kant  \cite{steffens1981, crowe1994history}, resonate in the history of mathematics. Despite their brief period of prominence, quaternions continue to be relevant today, demonstrating how a mathematical discovery can have a lasting impact and evolve over time.

In the 20th century, once eclipsed by vector analysis, quaternions experienced a significant revival, driven by their application in computer animation and programming  \cite{shoemake1985animating, vince2018quaternions}.

\section{Matrix representation of quaternions}

The development of matrix theory has roots that date back to the 2nd century BC, though it became more evident towards the end of the 17th century. One of the earliest indications of matrices comes from archaeological excavations in Babylon, where clay tablets were found that posed and solved problems involving linear equations. For example, a tablet from around 300 BC presents a problem involving calculating the size of two plots of land based on their grain production  \cite{hoyrup1994babylonian}.

The Chinese culture, between 200 BC and 100 BC, also approached matrices, as evidenced in the text ''Nine Chapters on the Mathematical Art'' \cite{shen1999nine}. This text contains problems that lead to systems of linear equations and shows a particular interest in their resolution. In one of the problems, the amount of cereal contained in different bundles is determined using a method that is essentially a primitive form of Gaussian elimination  \cite{martzloff2007history}.

In the 16th century, Girolamo Cardano provided a rule for solving systems of two linear equations, known as the Rule of Cramer for $2 \times 2$ \cite{boyer1968} systems, essentially Cramer's rule. This method led to the definition of determinants. The notion of determinant appeared almost simultaneously in Japan and Europe and was developed by mathematicians such as Maclaurin and Cramer  \cite{knobloch1994gauss}.

Gaussian Elimination, found for the first time in the ''Nine Chapters on the Mathematical Art'' \cite{shen1999nine} and attributed to Gauss, was used to solve systems of linear equations and was fundamental in his studies on the orbit of the asteroid Pallas. Subsequently, Cauchy introduced the term ''determinant'' and developed fundamental results in the field. Jacobi, Cayley, Eisenstein, and other mathematicians continued to advance matrix theory, developing significant notations and results.

The concept of a matrix was initially introduced in 1851 by the mathematician James Joseph Sylvester in his work titled ''On the relations between the minor determinants of linearly equivalent quadratic functions''  \cite{sylvester1851xxxvii}. In this work, Sylvester defined a matrix as a ''rectangular array of terms'', thus laying the fundamental groundwork for its study and application in various branches of mathematics and science.

Over the course of the historical evolution of quaternions, James Sylvester established contact with his colleague Arthur Cayley, whose contributions were fundamental in the algebraic aspect of matrices. Cayley introduced essential concepts such as zero and unit matrices, in addition to addressing matrix addition, highlighting its associative and commutative properties. In 1853, he published a note that marked the first appearance of the concept of a matrix inverse  \cite{cayley1854vii}. Later, in 1858, Sylvester, in his ''Memoir on the theory of matrices'', presented the first abstract definition of a matrix  \cite{cayley1858ii}, revealing that the arrays of coefficients previously studied in quadratic forms and linear transformations were exceptional cases of this broader concept  \cite{luzardo2006historia}.

Although quaternions were displaced by the vector analysis of Josiah Willard Gibbs and Oliver Heaviside in the 1880s, they experienced a resurgence in the late 20th century due to their efficacy in representing spatial rotations. They stood out in fields such as computer graphics, control theory, signal processing, and more  \cite{sweetser2005doing}.

The matrix representation of quaternions offers a valuable alternative perspective, and two matrix forms have been particularly relevant in this context.

\subsection{First fundamental form}
A structural homomorphism is established between quaternions and complex  $(2\times2)$ matrices through the correspondence rule  $f:\mathbb{H} \rightarrow M_2(\mathbb{C})$. In linear algebra, linear applications are also known as vector space homomorphisms due to their ability to preserve the operations and properties inherent to such spaces. A vector space homomorphism is a function that maintains the vector structure, that is, it preserves vector addition and scalar multiplication.

The correspondence is expressed as follows: 
\begin{equation*}
f(a+bi+cj+dk) = \begin{bmatrix}
a+bi & c+di \\
-c+di & a-bi\\
\end{bmatrix}
=A,
\end{equation*}
where, $a, b, c, d \in \mathbb{R}$ and $i, j, k$ are the unit imaginary elements of the quaternions. This relationship directly connects quaternions and complex matrices, preserving all algebraic operations.

The homomorphism  $f$ satisfies the following properties:

\begin{enumerate}
    \item 	$f(q_1 + q_2 )=f(q_1 )+f(q_2)$
    \item 	$f(r q_1)=r f(q_1)$
    \item  $f(q_1 q_2 )= f(q_1 ) f(q_2)$
    \item $f(0)=0$, where zero on the right side of the equality represents the zero complex matrix.
    \item 	$f(1)=1$, where 1 on the right side of the equality represents the identity complex matrix.
    \item 	The square of the norm of quaternion  $q_1$ is the determinant of its matrix  $A$.
    $$\mid q \mid ^2=a^2+b^2+c^2+d^2=det(A).$$
\end{enumerate}

The application of the homomorphism $f$ to the basis elements  ${1,i,j,k}$ $\in$ $\mathbb{H}$ produces the following matrices:
    \begin{equation*}
       f(1)=
        \begin{bmatrix}
        1 & 0\\
        0 & 1\\
        
        \end{bmatrix}=E_1,
      \hspace{1.5cm}
       f(i)=
        \begin{bmatrix}
        i & 0\\
        0 & -i\\
        
        \end{bmatrix}=I_1,
         \end{equation*}

    \begin{equation*}
         f(j)=
        \begin{bmatrix}
        0 & 1\\
        -1 & 0\\
        
        \end{bmatrix}=J_1,
        \hspace{1.5cm}
       f(k)=
        \begin{bmatrix}
        0 & i\\
        i & 0\\
        
        \end{bmatrix}=K_1.
    \end{equation*}

Matrix $A$ can be expressed as a linear combination of the previous matrices in the following manner:

 \begin{equation*}
       A=a
        \begin{bmatrix}
        1 & 0\\
        0 & 1\\
        
        \end{bmatrix}+b
        \begin{bmatrix}
        i & 0\\
        0 & -i\\
        
        \end{bmatrix}+c
        \begin{bmatrix}
        0 & 1\\
        -1 & 0\\
        
        \end{bmatrix}+d
        \begin{bmatrix}
        0 & i\\
        i & 0\\
        
        \end{bmatrix},
    \end{equation*}
that is, 
$$A=aE_1+bI_1+cJ_1+dK_1.$$

The properties of the homomorphism  $f$ establish a coherent connection between the operations and properties of quaternions and complex matrices. This relationship simplifies the study of quaternions by allowing the use of tools and techniques from matrix algebra, which proves beneficial in fields such as physics and engineering.

\subsection{Second fundamental form}

We define the algebra homomorphism from the quaternion algebra  $\mathbb{H}$ to the algebra of real  $(4\times4)$ matrices $M_4(\mathbb{R}$), with the following correspondence rule \cite{lin2022generalization}:
$$g:\mathbb{H} \rightarrow M_4(\mathbb{R}),$$ such that:
\begin{equation*}
       g(q)=g(a+bi+cj+dk)=
        \begin{bmatrix}
        a & b & c & d\\
        -b & a & -d & c\\
        -c & d & a & -b\\
        -d & -c & b & a\\
        
        \end{bmatrix}
        =C.
\end{equation*}        

The homomorphism $g$ satisfies the following properties:
\begin{enumerate}
    \item $g(q_1 + q_2 )=g(q_1 )+g(q_2)$
    \item $g(r q_1)=r g(q_1)$
    \item $g(q_1 q_2 )= g(q_1 ) g(q_2)$
    \item $g(0)=0$, where the quaternion $0$ is mapped to the $0$ real matrix.
    \item $g(1)=I$, where the quaternion $1$ is mapped to the identity real matrix $I$.
    \item The square of the norm of the quaternion  $q_1$ is equal to the determinant of its matrix  $C$:
    $$\mid q \mid ^2=a^2+b^2+c^2+d^2=det(C).$$
\end{enumerate}
    
The application of the homomorphism  $g$ to the basis values  $\{1,i,j,k\}$ of $\mathbb{H}$, results in the matrices:
    \begin{equation*}
       g(1)=
        \begin{bmatrix}
        1 & 0 & 0 & 0\\
        0 & 1 & 0 & 0\\
        0 & 0 & 1 & 0\\
        0 & 0 & 0 & 1\\
        
        \end{bmatrix}=E_2,
      \hspace{1.5cm}
       g(i)=
        \begin{bmatrix}
         0 & 1 & 0 & 0\\
        -1 & 0 & 0 & 0\\
        0 & 0 & 0 & -1\\
        0 & 0 & 1 & 0\\
        
        \end{bmatrix}=I_2,
         \end{equation*}

    \begin{equation*}
         g(j)=
        \begin{bmatrix}
        0 & 0 & 1 & 0\\
        0 & 0 & 0 & 1\\
        -1 & 0 & 0 & 0\\
        0 & -1 & 0 & 0\\
        
        \end{bmatrix}=J_2,
        \hspace{1.5cm}
       g(k)=
        \begin{bmatrix}
        0 & 0 & 0 & 1\\
        0 & 0 & -1 & 0\\
        0 & 1 & 0 & 0\\
        -1 & 0 & 0 & 0\\
        
        \end{bmatrix}=K_2.
    \end{equation*}

Matrix $C$ can be represented as a linear combination of the previous matrices in the following manner:
 \begin{equation*}
       C=a
        \begin{bmatrix}
        1 & 0 & 0 & 0\\
        0 & 1 & 0 & 0\\
        0 & 0 & 1 & 0\\
        0 & 0 & 0 & 1\\
        
        \end{bmatrix}+b
        \begin{bmatrix}
        0 & 1 & 0 & 0\\
        -1 & 0 & 0 & 0\\
        0 & 0 & 0 & -1\\
        0 & 0 & 1 & 0\\
        
        \end{bmatrix}+c
        \begin{bmatrix}
        0 & 0 & 1 & 0\\
        0 & 0 & 0 & 1\\
        -1 & 0 & 0 & 0\\
        0 & -1 & 0 & 0\\
        
        \end{bmatrix}+d
        \begin{bmatrix}
        0 & 0 & 0 & 1\\
        0 & 0 & -1 & 0\\
        0 & 1 & 0 & 0\\
        -1 & 0 & 0 & 0\\
        
        \end{bmatrix},
    \end{equation*}
such that:
$$C=aE_2+bI_2+cJ_2+dK_2.$$    
The matrix forms of quaternions, through homomorphisms with complex and real matrices, have proven to be powerful tools in various disciplines. They facilitate the analysis and application of quaternions in fields as diverse as bioinformatics, quantum physics, and orbital mechanics. This resurgence and expansion of the relevance of quaternions underscore their continuing importance in scientific and technological research and development.

\section{Application of Quaternions}

Quaternions offer a robust mathematical framework for describing and analyzing geometric and physical phenomena in three-dimensional space. This section will explore some of the most prominent applications of quaternion. We will begin by examining the rotations of an object in three-dimensional space and their relationship with the quaternion group. Additionally, we will explore the role of quaternions in quantum physics, specifically in the representation of Pauli matrices, which play a crucial role in describing the spin states of subatomic particles. Through these applications, we can appreciate the versatility and utility of quaternions in various fields of science and engineering.

\subsection{Rotations in the Plane}

The section dedicated to rotations in the two-dimensional  $(2D)$, plane, both using the product of two complex numbers and the product of a matrix and a complex number, offers a deep understanding of how these fundamental geometric transformations are carried out. Considering rotations in the complex plane around the origin, the importance of complex numbers of unit length in representing these rotations is highlighted. The use of complex multiplication shows how a complex number can rotate another around the origin without altering its magnitude, thus revealing the essential relationship between rotations and complex numbers. An alternative perspective for understanding these transformations is illustrated by introducing $2 \times 2$ orthogonal matrices and their association with rotations in the plane. The relationship between rotations in the plane and the special orthogonal group $SO(2)$ is established, providing a solid foundation for studying rotations in more general contexts. Detailed examples demonstrate how complex numbers and matrices can perform rotations in the plane, offering a complete and rich view of this fundamental concept in geometry and linear algebra.

\subsubsection{Rotations in the Plane Using the Product of Two Complex Numbers}

The rotation of a point $z$ in the Argand diagram  $(\mathbb{R}^2)$ or the two-dimensional complex space $(\mathbb{C})$ is performed around a specific point, which acts as the pivotal axis of rotation. The main rotations occur around the origin, while rotations around any other arbitrary point are called general rotations. The rotation angle and the rotation point (origin or pivot) around which the transformation will take place must be specified to carry out a rotation. A positive rotation angle indicates a counterclockwise rotation from the positive real axis toward the positive imaginary axis. In contrast, a negative rotation angle results in a clockwise rotation from the positive real axis toward the negative imaginary axis. When the rotation point is at the origin, the distance $r = |z|$ represents the magnitude of the complex number $z$ with respect to the origin. The action angle coincides with the angle formed by the corresponding complex number, considered as a vector in $(\mathbb{R}^2)$ with the positive real axis. Complex numbers of unit length can represent rotations in the complex plane around the origin. The description of these rotations is simplified using complex multiplication, leading us to a straightforward formula for describing these transformations, as detailed below.

Using the definition of polar representation and Euler's formula for a complex number  \cite{brown2009complex}, we consider the numbers  $z_1$ and $z_2$:

\begin{equation}
    z_1 = |z_1|e^{i\theta} = |z_1|(\cos{\theta} +i\sin{\theta}), \;\;\;\;\;\;\;\;  z_2 = |z_2|e^{i\alpha} = |z_2|(\cos{\alpha} +i\sin{\alpha}).
\end{equation}

When multiplying the two complex numbers, we obtain:

\begin{equation}
    z_1z_2 = (|z_1|e^{i\theta})(|z_2|e^{i\alpha}).
\end{equation}

If we consider that the complex  $z_1$ is unitary, that is, it has a norm of one ($|z_1| = 1$), then

\begin{equation}
z_1z_2 = (e^{i\theta})(|z_2|e^{i\alpha}) = |z_2|e^{i(\theta + \alpha)} = |z_2|(\cos(\theta + \alpha) + i\sin(\theta + \alpha))
\end{equation}

As we can see in the previous product, $z_1$ rotates $z_2$ by an angle $\theta$ around the origin without changing its magnitude.

\textbf{Examples.}

For example, consider the complex number $z_1=i$ purposefully for two reasons. First, this number is an element of the basis of the complex numbers $\mathbb{C}$. Second, in complex number theory, it can be interpreted as the generator of positive rotations in two dimensions. Then, consider the following two complex numbers:  
\begin{eqnarray}
    z_1 &=& (0,1) = 0+i = \cos(90^\circ) + i\sin(90^\circ) = e^{i(90^\circ)}, \nonumber \\
    z_2 &=& \left(\frac{\sqrt{3}}{2},\frac{1}{2}\right) = \frac{\sqrt{3}}{2} + \frac{1}{2}i = \cos(30^\circ) + i\sin(30^\circ) = e^{i(30^\circ)}.
\end{eqnarray}

\begin{figure}[h]
    \centering
    \includegraphics[width=7cm, height=6cm]{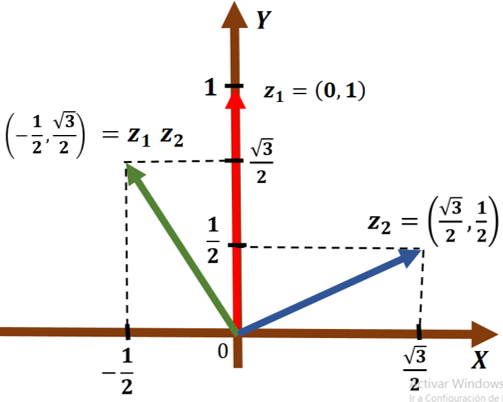}
    \hspace{1.3cm}
    \includegraphics[width=6cm, height=6cm]{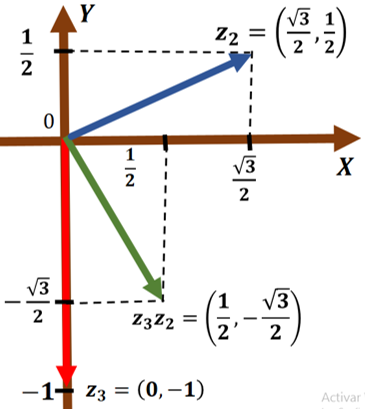}
    \caption{The left panel of this figure shows the positive rotation of two complex numbers, $z_1z_2$, while the right panel shows the negative rotation of two other complex numbers, $z_3z_2$, both in their Cartesian representation in the Argand diagram. Source: Author's elaboration.}
    \label{rota}        
\end{figure}

First, let's consider the complex product in its Cartesian representation and sum  $z_1z_2$:
\begin{align*}
    z_1z_2 &= (0,1)\left(\frac{\sqrt{3}}{2},\frac{1}{2}\right) = (0+i)\left(\frac{\sqrt{3}}{2}+\frac{1}{2}i\right) \\
    &= \left(0-\frac{1}{2}\right) + \left(0+\frac{\sqrt{3}}{2}\right)i \\
    &= -\frac{1}{2} + \frac{\sqrt{3}}{2}i = \left(-\frac{1}{2},\frac{\sqrt{3}}{2}\right).
\end{align*}
We observe that the product  $z_1z_2$ produces a positive rotation of $z_2$ by an angle of $90^\circ$ counterclockwise around the origin, as illustrated in the left panel of Figure \ref{rota}. Similarly, using its polar representation and Euler's formula, we obtain:
\begin{align*}
    z_1z_2 &= (\cos(90^\circ) + i\sin(90^\circ))(\cos(30^\circ) + i\sin(30^\circ)) \\
    &= e^{i(90^\circ)} \cdot e^{i(30^\circ)} \\
    &= e^{i(120^\circ)} \\
    &= \cos(120^\circ) + i\sin(120^\circ).
\end{align*}
This result also produces a rotation of  $z_1$ to $z_2$ by an angle of $90^\circ$ counterclockwise around the origin, as in the previous case.
        
If we now consider $z_3=-i$, we proceed to calculate the product  $z_3z_2$ using its polar representation and Euler’s formula:
\begin{align*}
    z_3z_2 &= (\cos(-90^{\circ}) + i\sin(-90^{\circ}))(\cos(30^{\circ}) + i\sin(30^{\circ})) \\
    &= e^{i(-90^{\circ})} \cdot e^{i(30^{\circ})} = e^{i(-60^{\circ})} \\
    &= \cos(-60^{\circ}) + i\sin(-60^{\circ})
\end{align*}
Notice that in this product, $z_3$ rotates $z_2$ by an angle of $-90^{\circ}$ clockwise around the origin (negative direction), as shown in the right panel of Figure \ref{rota}.

In another example, we explore the multiplication of complex numbers and how this operation relates to rotation in the complex plane while preserving the norm. We start by considering two complex numbers: 
\begin{equation}
    z_2=\left(\frac{\sqrt{3}}{2},\frac{1}{2}\right)=\frac{\sqrt{3}}{2}+\frac{1}{2}i, \;\;\;\;\;\;\;\;\;\; z_4= (0,4)= 0+4i=4e^{i(90^\circ)}
\end{equation}

Notice that $z_2$ is unitary:
\begin{equation}
    |z_2|=\sqrt{\left(\frac{\sqrt{3}}{2}\right)^2+\left(\frac{1}{2}\right)^2}=\sqrt{\frac{4}{4}}=1,
\end{equation} 
while the norm of $|z_4|=4$.

\begin{figure}[h]
       \centering
       \includegraphics[width=6cm, height=6cm]{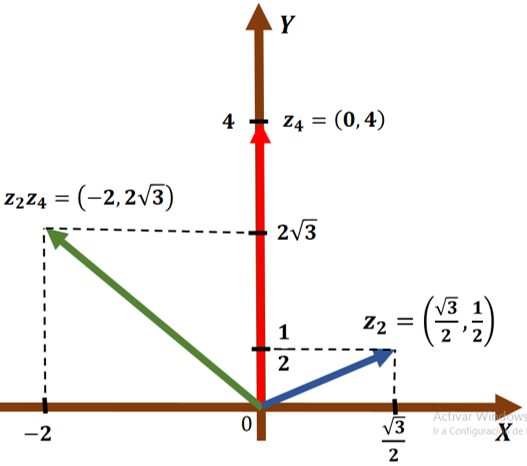}
       \hspace{1cm}
        \includegraphics[width=6cm, height=6cm]{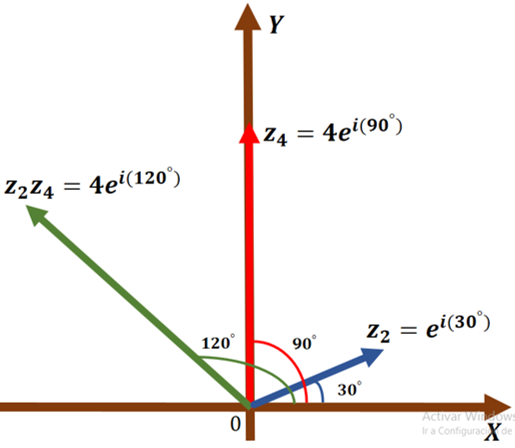}
        \caption{Left panel: Positive rotation $z_2z_4$ os the complex product in its cartesian representation. Right panel: same product but in its polar form on Argand's diagram. Source: Author's elaboration.}
        \label{rotación}        
    \end{figure}

Using the Cartesian representation and sum of two terms, we multiply these complex numbers:
\begin{align*}
    z_2z_4 &= \left(\frac{\sqrt{3}}{2},\frac{1}{2}\right)(0,4)=\left(\frac{\sqrt{3}}{2}+\frac{1}{2}i\right)(0+4i)\\
    &= (0-2)+(2\sqrt{3}+0)i\\
    &= -2+2\sqrt{3}i\\
    &= (-2,2\sqrt{3}).
\end{align*}
Notice that in the product, $z_2$ rotates $z_4$ by an angle of $30^\circ$ counterclockwise around the origin. Furthermore, the norm of the product $|z_2z_4|=\sqrt{(-2)^2 +(2\sqrt{3})^2}=\sqrt{16}=4$, is the same as that of $|z_2|$, i.e., the magnitude is preserved, as can be seen in the left panel of Figure  \ref{rotación}.

Now we perform the complex product in its polar representation and Euler’s formula: 
\begin{align*}
    z_2z_4 &= (\cos(30^\circ)+i\sin(30^\circ))4(\cos(90^\circ)+i\sin(90^\circ))\\
    &= e^{i(30^\circ)}  4e^{i(90^\circ)} \\
    &= 4e^{i(120^\circ)}\\
    &= 4(\cos(120^\circ)+i\sin(120^\circ))
\end{align*}
Notice that in the product, $ z_2$ rotates $z_4$ by an angle of $30^\circ$ counterclockwise around the origin, as shown in the right panel of Figure \ref{rotación}.

In general, for all complex numbers \( z = e^{i\theta} = \cos{\theta} + i\sin{\theta} = a + ib \) of unit norm (\( |z| = 1 \)), we can define the set:
\begin{equation}
    S^1 = \{ z \in \mathbb{C} : a^2 + b^2 = \cos^2{\theta} + \sin^2{\theta} = 1 
\end{equation}

The set $S^1 $ forms a multiplicative group, meaning the product of two complex numbers of unit norm is again a complex number of unit norm. The product of complex numbers has the property of being associative: $z_1(z_2z_3) = (z_1z_2)z_3$. As the complexes in this set are all non-zero, they have a multiplicative inverse $z^{-1} = \overline{z} = \cos{\theta} - i\sin{\theta} = a - ib$. Finally, the complex  $z = 1 + i0 = 1$ is the group's identity. Given that rotations are defined by the product $z_1z_2$ with $|z_1| = 1$, it is then said that the group  $S^1$ describes two-dimensional (2D) rotations in the Argand diagram $\mathbb{R}^2$ or the complex space $\mathbb{C}$.

\subsubsection{Rotations in the Plane Using the Product of a Matrix and a Complex Number}

In the previous subsection, we observed that two-dimensional ($2D$) rotations are described by the group  $S^1$. This subsection relates the group $S^1$ to $2 \times 2$ matrices with real entries. 

Consider the function $g: S^1 \rightarrow M_2(\mathbb{R})$ defined by:
\begin{equation*}
    g(z) = \begin{bmatrix}
        \cos{\theta} & -\sin{\theta} \\
        \sin{\theta} & \cos{\theta} \\
    \end{bmatrix} = \begin{bmatrix}
        a & -b \\
        b & a \\
    \end{bmatrix} = A,
\end{equation*}
for every $z = (a,b) = e^{i\theta} = \cos{\theta} + i\sin{\theta} = a + ib \in S^1$. 

The matrix $A$ can be rewritten as a linear combination in the following way:
\begin{equation*}
    A = I \cos(\theta) + i \sin(\theta), 
\end{equation*}
where  
\begin{equation}
    I = \begin{bmatrix}
        1 & 0 \\
        0 & 1 \\
    \end{bmatrix}, \quad
    i = \begin{bmatrix}
        0 & -1 \\
        1 & 0 \\
    \end{bmatrix}, \label{Matrices}
\end{equation}

and

\begin{equation*}
    A = \begin{bmatrix}
        \cos{\theta} & -\sin{\theta} \\
        \sin{\theta} & \cos{\theta} \\
    \end{bmatrix} = \begin{bmatrix}
        1 & 0 \\
        0 & 1 \\
    \end{bmatrix} \cos(\theta) + \begin{bmatrix}
        0 & -1 \\
        1 & 0 \\
    \end{bmatrix} \sin(\theta).
\end{equation*}

The multiplication table presented in Table \ref{tab:rota} can be used to combine the operations of rotation.

\begin{table}[h]
    \centering
    \caption{Multiplication table representing the product of matrices as described in equation (\ref{Matrices}).}
    \begin{tabular}{ r | l | c }
      & $I$ & $i$ \\ \hline
    $I$ & $I$ & $i$ \\
    $i$ & $i$ & $-I$ \\ 
    \end{tabular}
    \label{tab:rota}
\end{table}

The algebraic operations in Table \ref{tab:rota} are isomorphic to the system of operations in complex numbers ($\mathbb{C}$), where $I$ represents the number $1$ and $i$ represents the imaginary unit $\sqrt{-1}$. Therefore, rotations can be defined both by the multiplication of complex numbers (as demonstrated in the previous subsection) and by the multiplication of matrices.

In this way, the function $g$ fulfills the following conditions:
\begin{enumerate}
    \item The function $ g$ preserves operations, that is, $ g (z_1z_2) = g (z_1) g (z_2) $. This means that the product on the left side of the equality is the product of the complex numbers, and the product on the right side of the equality is the product of the corresponding matrices.
    
    \item The determinant of the matrix  $A$ is related to the following equality: $$\det(A)=a^2+b^2=\cos^2{\theta}+\sin^2{\theta}=1$$
\end{enumerate}

The set of matrices that meet these characteristics is called the special orthogonal matrices  $SO(2)$. This set $SO(2)$ forms a multiplicative group, which means that the product of two orthogonal matrices is again an orthogonal matrix. Additionally, the matrix product has the property of being associative, that is, $A(BC)=(AB)C$. Since the matrices in this set are all non-zero and have a determinant equal to one, they have an inverse given by:

\begin{equation*}
    A^{-1}=
    \begin{bmatrix}
        a & b\\
        -b & a\\
    \end{bmatrix}.
\end{equation*}

The inverse matrix has the following property: 
\begin{equation*}
    AA^{-1}=
    \begin{bmatrix}
        a & -b\\
        b & a\\
    \end{bmatrix}
    \begin{bmatrix}
        a & b\\
        -b & a\\
    \end{bmatrix}
    =
    \begin{bmatrix}
        a^{2}+b^{2} & ab-ba\\
        ba-ab & b^{2}+a^{2}\\
    \end{bmatrix}
    =
    \begin{bmatrix}
        1 & 0\\
        0 & 1\\
    \end{bmatrix}
    = A^{-1}A.
\end{equation*}

Lastly, the identity matrix  $I$ (\ref{Matrices}) represents the group's identity. Consequently, the set $SO(2)$ is known as the special orthogonal group.

Rotations in the complex plane $\mathbb{C}$ or in two-dimensional space $\mathbb{R}^2$ are defined by the product of the matrix  $A \in SO(2)$ with the complex number $z=
    \begin{bmatrix}
        x \\
        y\\
    \end{bmatrix} \in \mathbb{R}^2$ (viewed as a column vector), as shown below:

\begin{equation*}
    Az=
    \begin{bmatrix}
        a & -b\\
        b & a\\
    \end{bmatrix}
    \begin{bmatrix}
        x \\
        y\\
    \end{bmatrix}
    =
    \begin{bmatrix}
        ax-by\\
        bx+ay \\
    \end{bmatrix}.
\end{equation*}
Since rotations are defined by the product $Az$, it is said that the group of orthogonal matrices $SO(2)$ describes the rotations in two dimensions $(2D)$.

\textbf{Examples:}

The following examples will verify how positive and negative rotations can be performed using a matrix $A \in SO(2)$ on a vector $z \in \mathbb{R}^2$.

\begin{figure}[h]
       \centering
       \includegraphics[width=6cm, height=6cm]{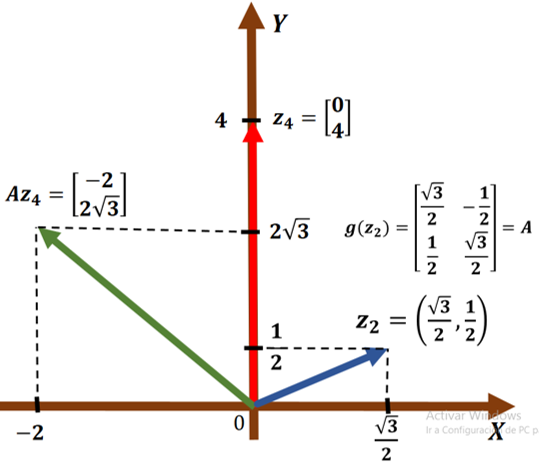}
       \hspace{1cm}
        \includegraphics[width=6cm, height=6cm]{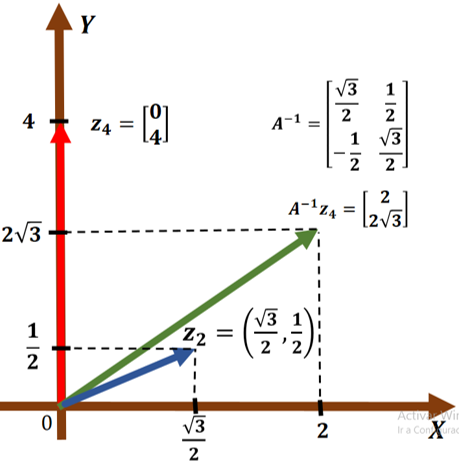}
        \caption{The left panel shows the positive rotation $Az_4$ and the right panel shows the negative rotation $A^{-1}z_4$ of the product of a matrix and the vector $z_4$. Source: Author's elaboration.}
        \label{matrizrota}        
    \end{figure}

\textbf{Positive Rotation } of a matrix $A \in SO(2)$ over a vector  $z \in \mathbb{R}^2$. We will use the same complex numbers $z_2=(\frac{\sqrt{3}}{2},\frac{1}{2})$ and $z_4=(0,4)$ from the example in the previous subsection. Applying the function $g$ to the complex number $z_2$, we obtain:
    
    \begin{equation*}
       g(z_2)=
        \begin{bmatrix}
        \frac{\sqrt{3}}{2} & -\frac{1}{2} \\
        \frac{1}{2} & \frac{\sqrt{3}}{2}\\
        
        \end{bmatrix}
        =A.
    \end{equation*}
Then, applying the product of the matrix  $A$ with the vector $z_4$ viewed as a column vector:
    \begin{equation*}
       z_4=
       \begin{bmatrix}
         0 \\
        4\\
        \end{bmatrix},
    \end{equation*}
then, we obtain:
    \begin{equation*}
       Az_4=
        \begin{bmatrix}
        \frac{\sqrt{3}}{2} & -\frac{1}{2} \\
        \frac{1}{2} & \frac{\sqrt{3}}{2}\\
        
        \end{bmatrix}
        \begin{bmatrix}
         0 \\
        4\\
        
        \end{bmatrix}
        =
         \begin{bmatrix}
         0-2 \\
        0+2\sqrt{3}\\
        
        \end{bmatrix}
        =
         \begin{bmatrix}
         -2 \\
        2\sqrt{3}\\
        
        \end{bmatrix}.
    \end{equation*}
This expression implies that the matrix  $A$ rotates the vector $z_4$ by an angle of $30^{\circ}$ counterclockwise around the origin, as shown in the left panel of Figure \ref{matrizrota}.
    
\textbf{Negative rotation} of a matrix $A \in SO(2)$ over a vector $z \in \mathbb{R}^2$. We will use the same complex number $z_2=(\frac{\sqrt{3}}{2},\frac{1}{2})$ y $z_4=(0,4)$ from the example in the previous subsection, but this time considering its inverse matrix $A^{-1}$: 
\begin{equation*}
    A^{-1}=
    \begin{bmatrix}
    \frac{\sqrt{3}}{2} & \frac{1}{2} \\
        -\frac{1}{2} & \frac{\sqrt{3}}{2}\\
    \end{bmatrix}
\end{equation*}
    
Now, applying the product of the matrix $A^{-1}$ with the vector $z_4$ viewed as a column vector:
\begin{equation*}
    z_4=
       \begin{bmatrix}
         0 \\
        4\\
        \end{bmatrix},
\end{equation*}
then, we can obtain the following:
\begin{equation*}
    A^{-1}z_4=
    \begin{bmatrix}
        \frac{\sqrt{3}}{2} & \frac{1}{2} \\
        -\frac{1}{2} & \frac{\sqrt{3}}{2}\\
        
    \end{bmatrix}
    \begin{bmatrix}
        0 \\
        4\\
    \end{bmatrix}
    =
    \begin{bmatrix}
        0+2 \\
        0+2\sqrt{3}\\
    \end{bmatrix}
    =
    \begin{bmatrix}
        2 \\
        2\sqrt{3}\\
    \end{bmatrix}.
\end{equation*}
    This expression implies that the matrix  $A^{-1}$ rotates the vector $z_4$ by an angle of $-30^{\circ}$ clockwise around the origin, as shown in the right panel of Figure \ref{matrizrota}.
    
\subsubsection{Rotations in space} 

Hamilton spent years developing an algebra of rotations in $\mathbb{R}^3$ using ordered triples of real numbers \cite{crowe1969history, crowe1994history}. However, one day, he realized he could achieve his goal using ordered quaternions of real numbers.

Initially \cite{lyons2003elementary}, note that a rotation around the origin in $\mathbb{R}^3$ can be specified by a vector representing the axis of rotation and a rotation angle around that axis. We adopt the convention that rotations are performed counterclockwise for positive angles and clockwise for negative angles, viewed from the tip of the vector.

Specifying a rotation using an axis vector and an angle is not unique. The rotation determined by the vector $v$ and the angle $\theta$ is equivalent to the rotation defined by the pair $(kv,\theta + 2n\pi)$, where $k$ is any positive scalar and  $n$ is any integer. The pair $(-v,-\theta)$ also defines the same rotation. Using linear algebra, this is achieved using the nine entries of a $3 \times 3$ orthogonal matrix. However, there is a more efficient method for working with quaternions and performing practical calculations to obtain rotations.

Four real numbers are sufficient to specify a rotation: three coordinates for a vector and one real number to indicate the angle. Hereafter, the algebraic solution to this fact is outlined.

Defining a formula for rotations in $\mathbb{R}^3$ does not simply involve multiplying two quaternions, as multiplying an $\mathbb{R}^3$ vector with a quaternion whose real part is non-zero results in a product with a non-zero real part. Thus, it will not be a vector in $\mathbb{R}^3$. The problem lies in eliminating the real part so that the product is a vector in $\mathbb{R}^3$.

We observe that a vector $v \in \mathbb{R}^3$ is a pure quaternion whose real part is zero. Consider a unit quaternion $q = q_0 + q_1i + q_2j + q_3k = q_0 + \bold{q}$, where its norm or magnitude $q_0^2 + \|\bold{q}\|^2 = 1$, lo implying the existence of an angle $\theta$ such that $cos^2\theta = q_0^2$ and $sin^2\theta = \|\bold{q}\|^2$. Indeed, there is a unique $\theta \in [0,\pi]$ such that $cos\theta = q_0$ and $sin\theta = \|\bold{q}\|$. 

The unit quaternion can now be expressed in terms of the angle $\theta$ and the unit vector  $\bold{u} = \frac{\bold{q}}{\|\bold{q}\|}$ as follows:
$$q = \cos\theta + \bold{u}\sin\theta.$$

We define an operator on vectors  $v \in \mathbb{R}^3$ using the quaternion product and the unit form of the quaternion  $q$. The following equalities are met:
\begin{align}
L_q(v) &= qvq^{\ast} \nonumber \\
       &= (q_0^2 - \|\bold{q}\|^2)v + 2(\bold{q} \cdot v)\bold{q} + 2q_0(\bold{q} \times v) \nonumber \\
       &= \cos\theta \cdot v + (1 - \cos\theta)(\bold{u} \cdot v)\bold{u} + \sin\theta \cdot (\bold{u} \times v). \label{Lqv}
\end{align}

Two observations about the operator: first, the operator does not alter the length of the vector  $v$, as:
$$\|L_q(v)\| = \|qvq^{\ast}\| = |q| \cdot \|v\| \cdot |q^{\ast}| = \|v\|$$

The second observation is that the operator  $L_q$ does not change the direction of the vector $v$ if it lies along $\bold{q}$. To verify this, consider $v = k\bold{q}$, then:
$$qvq^{\ast} = q(k\bold{q})q^{\ast} = (q_0^2 - \|\bold{q}\|^2)(k\bold{q}) + 2(\bold{q} \cdot k\bold{q})\bold{q} + 2q_0(\bold{q} \times k\bold{q}) = k(q_0^2 + \|\bold{q}\|^2)\bold{q} = k\bold{q}$$

These observations suggest that the operator $L_q$ acts as a rotation around  $q$. Before starting this formally, we note that the operator  $L_q$ is linear over $\mathbb{R}^3$. For any two vectors $v_1, v_2 \in \mathbb{R}^3$ and scalars $a_1, a_2 \in \mathbb{R}$, it is true that:
$$L_q(a_1v_1 + a_2v_2) = a_1L_q(v_1) + a_2L_q(v_2)$$

With these observations about $L_q$, we can affirm the following result. For any unit quaternion of the form:
\begin{equation}
    q = q_0 + \bold{q} = \cos\frac{\theta}{2} + u\sin\frac{\theta}{2}, \label{Eqq}
\end{equation} and any vector $v \in \mathbb{R}^3$, the action of the operator $$L_q(v) = qvq^{\ast}$$ over $v$ is equivalent to a rotation of an angle $\theta$ of the vector around  $u$ as rotation axis. For a formal demonstration of this fact, see \cite{jia2008quaternions}.

\textbf{Examples:}

\textbf{a. Rotation of a vector in $\mathbb{R}^3$ under the operator $L_q$.}

Consider a rotation around an axis defined by the vector  $(1,1,1) \in \mathbb{R}^3$ at an angle of $\frac{2\pi}{3}$. Along this axis, the canonical vectors $i$, $j$ y $k$ generate the same cone when rotated by an angle of  $2\pi$.

We define the unit vector  $u = \frac{1}{\sqrt{3}}(1,1,1)$ and consider the rotation angle $\theta = \frac{2\pi}{3}$. The quaternion $q$ that defines the rotation is given by (\ref{Eqq}):
\begin{equation}
    q = \cos\frac{\theta}{2} + u\sin\frac{\theta}{2}.
\end{equation}

Substituting the value of $u$ and $\theta = \frac{2\pi}{3}$ into the formula for  $q$, we obtain:
\begin{equation}
    q = \cos\left(\frac{\frac{2\pi}{3}}{2}\right) + \frac{1}{\sqrt{3}}(1,1,1) \sin\left(\frac{\frac{2\pi}{3}}{2}\right) = \frac{1}{2} + \frac{1}{\sqrt{3}}(1,1,1)\frac{\sqrt{3}}{2}= \frac{1}{2} + \frac{1}{2}i + \frac{1}{2}j + \frac{1}{2}k
\end{equation}

Note that the quaternion $q$, as defined, is unitary, meaning $|q| = 1$.

\textbf{b. Rotation of a vector $v=i$ under the operator $L_q$.}

Let's calculate the rotation effect on the canonical vector of the basis of  $\mathbb{R}^3$, $v=i=(1,0,0)$. In other words, we want to determine  $L_q(v)=qvq^{\ast}$. We will apply the second definition of quaternion product twice to do this. First, we calculate the product $qv$ using the equation (\ref{2daPC}). Considering the values $q=\frac{1}{2}+\frac{1}{2}i + \frac{1}{2}j+\frac{1}{2}k$ y $v=0+i+0j+0k$, we obtain:

\begin{equation}
    qv=-\frac{1}{2}+\frac{1}{2}i+\frac{1}{2}j-\frac{1}{2}k.
\end{equation}

Now, to calculate the product  $(qv)q^{\ast}=qvq^{\ast}$, we apply again the second definition of quaternion product  (\ref{2daPC}), using $qv=-\frac{1}{2}+\frac{1}{2}i+\frac{1}{2}j-\frac{1}{2}k$ y $q^{\ast}=\frac{1}{2}-\frac{1}{2}i-\frac{1}{2}j-\frac{1}{2}k$. The result obtained is as follows:

\begin{equation}
    qvq^{\ast}=\frac{1}{2}i+\frac{1}{2}j-\frac{1}{2}i+\frac{1}{2}j=j.
\end{equation}

\begin{figure}[h]
\centering
\includegraphics[width=7cm, height=6cm]{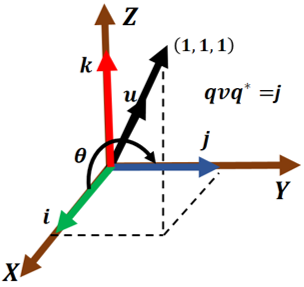}
\caption{The quaternion $q=\frac{1}{2}+\frac{1}{2}i + \frac{1}{2}j+\frac{1}{2}k$ rotates the vector $v=i \in \mathbb{R}^3$, via the operator $L_q(v)=qvq^\ast=j$ an angle of $\theta=\frac{2\pi}{3}$. Source: Author's elaboration.}
\label{rota3Di}
\end{figure}

Therefore, as expected, the unit quaternion $q$ rotates the basic vector  $v=i$ to the vector $j$. See Figure  \ref{rota3Di}.

\textbf{c. Rotation of the vector $v=j$ under the operator $L_q$.}

We will analyze the effect of rotation on the canonical vector of the base of $\mathbb{R}^3$, $v=j=(0,1,0)$, using the quaternion $q$ and the angle $\theta$ from the previous example. In this case, we apply equation (\ref{Lqv}) to the quaternions $q=\frac{1}{2}+\frac{1}{2}i + \frac{1}{2}j+\frac{1}{2}k$ y $v=0+0i+j+0k$, obtaining:
\begin{align}
    L_q(v) &= qvq^\ast = k.
\end{align}

\begin{figure}[h]
         \centering
         \includegraphics[width=7cm, height=6cm]{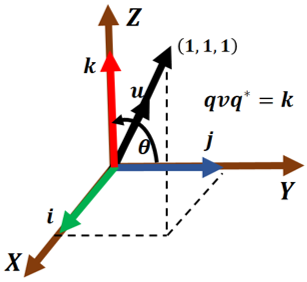}
         \caption{Quaternion $q=\frac{1}{2}+\frac{1}{2}i + \frac{1}{2}j+\frac{1}{2}k$ rotates vector $v=j \in \mathbb{R}^3$, via the operator $L_q(v)=qvq^\ast=j$ by an angle of $\theta=\frac{2\pi}{3}$. Source: Author's elaboration.}
         \label{rota3Dj}        
         \end{figure}

As expected, the unit quaternion $q$ rotates the basic vector  $v=j$, taking it to the vector $k$. This fact is illustrated in Figure  \ref{rota3Dj}.

\textbf{d. Rotation of vector $v=k$ using the quaternion $q$ and angle $\theta$.}

We'll examine the effect of rotation on the canonical vector of the base of  $\mathbb{R}^3$, $v=k=(0,0,1)$, using the quaternion $q$ and the angle  $\theta$. Essentially, we evaluate $L_q(v)=qvq^{\ast}$. For this case, we apply the second equation  (\ref{Lqv}) to $u=\frac{1}{\sqrt{3}}(1,1,1)$ and $\theta =\frac{2\pi}{3}$, obtaining:
\begin{equation}
    L_q(v)=qvq^\ast=i
\end{equation}
                
\begin{figure}[h]
         \centering
         \includegraphics[width=7cm, height=6cm]{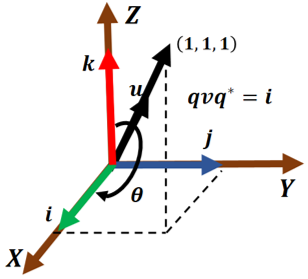}
         \caption{The quaternion $q=\frac{1}{2}+\frac{1}{2}i + \frac{1}{2}j+\frac{1}{2}k$ rotates the vector $v=k \in \mathbb{R}^3$, via the operator $L_q(v)=qvq^\ast=i$ an angle of $\theta=\frac{2\pi}{3}$. Source: Author's elaboration.}
         \label{rota3Dk}        
         \end{figure}

This result confirms our expectations: the unit quaternion $q$ rotates the basic vector $v=k$, taking it to the vector $i$, just as we anticipated, as shown in Figure \ref{rota3Dk} for a visual representation.

The rotation of the vector $v$ under the operator $L_q$ can also be interpreted from the perspective of an observer attached to the vector $v$. From this viewpoint, the observer perceives that the coordinate frame rotates through an angle $-\theta$ around the same axis defined by the quaternion $q$. Formally, this phenomenon is described as follows: for any unit quaternion $q=a+\bold{q}=cos \frac{\theta}{2}+usen\frac{\theta}{2}$ and any vector $v\in \mathbb{R}^3$, the action of the operator: 
\begin{equation}
    L_{q^{\ast}}=q^{\ast}v(q^{\ast})^{\ast}=q^{\ast}vq,
\end{equation}
acting on $v$ is equivalent to a rotation of the coordinate frame around the axis $u$ through an angle, while the vector $v$ remains unchanged. Conversely, the operator  $L_{q^{\ast}}$ rotates the vector $v$ with respect to the coordinate frame of the quaternion $q$ by an angle  $-\theta$.

On the other hand, from a physical viewpoint, the quaternionic operator $L_q(v)=qvq^{\ast}$ can be interpreted as a rotation of a point or vector with respect to a fixed coordinate frame. Meanwhile, the operator  $L_{q^{\ast}}=q^{\ast}vq$ can be understood as a rotation of the coordinate frame in relation to fixed space points.

By analogy with complex numbers  $(\mathbb{C})$, all quaternions $q=a+bi+cj+dk=\cos\frac{\theta}{2} +u\sin\frac{\theta}{2}$ of unit norm $|q|=1$ are defined as the set:

\begin{equation}
    S^3=\{q\in\mathbb{H}: a^2+b^2+c^2+d^2=1\} 
\end{equation}

The set $S^3$ forms a multiplicative group, meaning that the product of two quaternions of unit norm has again a quaternion of unit norm hence this product has the property of being associative, $q_1(q_2q_3)=(q_1q_2)q_3$, and as the quaternions in this set are all non-zero, they have a multiplicative inverse  $q^{-1}=q^{\ast}$. Lastly, the quaternion $q=1+0i+0j+0k=1$ is the identity of the group. Since rotations are defined by the operator $L_q$ with the quaternion $q \in S^3$, applied to any vector $v\in \mathbb{R}^3$, it is then said that the group $S^3$ describes rotations in three dimensions ($3D$) in space.

\subsubsection{Quaternions and Rotations of a Frame in $\mathbb{R}^3$}

In the context of this paper, we have explored various applications of quaternions in the study of three-dimensional rotations in $\mathbb{R}^3$. In this subsection, we focus on a particular case: the rotations of a three-dimensional frame. We begin by examining the group of quaternions and its algebraic structure, followed by an investigation into the Klein group and its role in the symmetries of the frame. Then, we explore how the operations of rotating the frame are linked to the operations of the Klein group. Subsequently, we investigate the connection between quaternions and the initial frame with strings tied (CIC), using the concept of the ''Dirac belt trick''. Finally, we present examples that illustrate how multiplications in the group of quaternions translate into specific rotations of the frame, providing a comprehensive view of the algebraic and geometric relationships involved.

\begin{flushleft}
\textbf{\textit{The quaternion group}}
\end{flushleft}

Considering the operation of the internal binary product of quaternions $\mathbb{H}$, the quaternion group is defined as the set:
\begin{equation*}
    \mathbb{Q}=\{1,i,j,k,-1,-i,-j,-k\},
\end{equation*}
which consists of 8 elements; thus, the multiplication table for the quaternion group, also known as the Cayley group table, represents the multiplication operations between the group elements. This table, shown in Table (\ref{multgrupo}), follows the order of conversion: row entry followed by column entry, and has been adapted by \cite{weissteinquaternion}. Considering the operation of the internal binary product of quaternions with their relationships in the basis  $\{1,i,j,k\}$.

\begin{table}[h]
\centering
\caption{Multiplication Table of the Quaternion Group $\mathbb{Q}$.}
\begin{tabular}{|c||c|c|c|c|c|c|c|c|}
\hline
$\cdot$ & $1$ & $i$ & $j$ & $k$ & $-1$ & $-i$ & $-j$ & $-k$ \\ \hline \hline
$1$ & $1$ & $i$ & $j$ & $k$ & $-1$ & $-i$ & $-j$ & $-k$ \\ \hline
$i$ & $i$ & $-1$ & $k$ & $-j$ & $-i$ & $1$ & $-k$ & $j$ \\ \hline
$j$ & $j$ & $-k$ & $-1$ & $i$ & $-j$ & $k$ & $1$ & $-i$ \\ \hline
$k$ & $k$ & $j$ & $-i$ & $-1$ & $-k$ & $-j$ & $i$ & $1$ \\ \hline
$-1$ & $-1$ & $-i$ & $-j$ & $-k$ & $1$ & $i$ & $j$ & $k$ \\ \hline
$-i$ & $-i$ & $1$ & $-k$ & $j$ & $i$ & $1$ & $k$ & $-j$ \\ \hline
$-j$ & $-j$ & $k$ & $1$ & $-i$ & $j$ & $-k$ & $-1$ & $i$ \\ \hline
$-k$ & $-k$ & $-j$ & $i$ & $1$ & $k$ & $j$ & $-i$ & $-1$ \\ \hline
\end{tabular}
\label{multgrupo}
\end{table}

The group multiplication can also be determined by the rules:  $i^2=j^2=k^2=-1$, $ij=k$, $jk=i$, $ki=j$, $ik=-j$, $ji=-k$, $kj=-i$, and the identity $1$. These rules are the same as in the quaternions $\mathbb{H}$ and define entirely the group multiplication.

\begin{flushleft}
\textbf{\textit{The Klein Group}}
\end{flushleft}
In group theory, the Klein group $V$, also known as the Klein four-group or \textit{Vierergruppe} in German, is named after the mathematician Felix Klein. This abelian group consists of four elements, each with its inverse. The group can be represented in various ways, with the most general form being: $V= \{e,a,b,ab\}$, where $e$ is the identity and $a$, $b$ are two distinct elements that generate the fourth element $ab$ when multiplied together. Each element in the set equals its own inverse: $a=a^{-1}$, $b=b^{-1}$ y $ab=(ab)^{-1}$. Additionally, it follows that $ab=(ab)^{-1}=b^{-1} a^{-1}=ba$. Thus, the set is closed under the operation of the product, and this operation is associative, which establishes that $V$ is a group. The Klein group is defined as the pair  $(V,\ast)$, where $V= \{e,a,b,c\}$ y $ab=c$, and its internal binary operation  $(\ast)$ is defined in Table \ref{Kleinpro} \cite{fraleigh1990linear}.

\begin{table}[h]
\centering
\caption{Multiplication Table of the Klein Group.}
\begin{tabular}{|c||c|c|c|c|}
\hline
$\ast$ & $e$ & $a$ & $b$ & $ab$ \\ \hline \hline
$e$    & $e$ & $a$ & $b$ & $ab$ \\ \hline
$a$    & $a$ & $e$ & $ab$ & $b$ \\ \hline
$b$    & $b$ & $ab$ & $e$ & $a$ \\ \hline
$ab$   & $ab$ & $b$ & $a$ & $e$ \\ \hline
\end{tabular}
\label{Kleinpro}
\end{table}

Its structure and properties make it an object of interest in group theory. This paper will see it as the symmetry group of a square in $\mathbb{R}^3$. This claim is explored in the following section. The book \cite{alexandroff2012introduction} is recommended for more detailed information on this topic.

\begin{flushleft}
\textbf{\textit{The Klein Group and the Initial Frame $(CI)$}}
\end{flushleft}

The rotation movement of a frame in $\mathbb{R}^3$ can be associated with an operation of the Klein group. For this, two fundamental steps are required, which are explained below.

\begin{enumerate}
    \item \textit{First step:} Consider a painting on a frame (yellow flower) labeled with the word ''Flower'' in the following order. In the top left corner is the letter  $F$, in the top right corner is the letter $l$, in the bottom right corner is the letter $o$, and finally, in the bottom left corner is the letter $r$. We define this frame as the initial or front of the frame  $(CI)$. On the back of the frame (rear side), consider a second image (blue flower) to differentiate it from the front of the frame; both frames are shown in Figure (\ref{cuadroinic}).
    
    \begin{figure}[h]
        \centering
        \includegraphics[width=5cm, height=5cm]{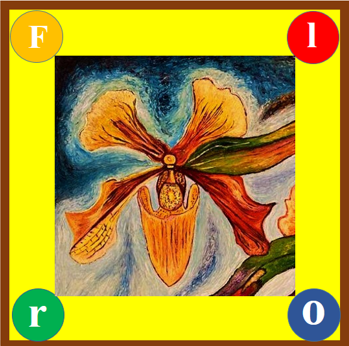}
        \hspace{1cm}
        \includegraphics[width=5cm, height=5cm]{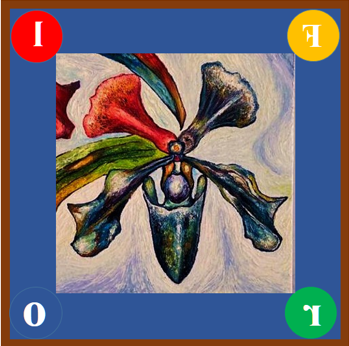}
        \caption{: The front and back of the initial frame $(CI)$, respectively. The frame is an original oil on canvas painted by the first author of this work. }
        \label{cuadroinic}        
    \end{figure}
    
    \item \textit{Second step:} We identify the elements of the Klein group with the rotations of the $(CI)$ in the following order. The identity element $(e)$ corresponds to the initial frame $(CI)$ and we call it the initial position \textit{posición inicial} $I$ of the frame. Element $(a)$ is associated with the rotation right or left of the frame from the initial position by a value of $\pi$, around the axis perpendicular to the plane of the frame, passing through the center of the frame (the upside-down yellow flower). We call this rotation \textit{rotación} $R_2$. The element $(b)$ is related to the rotation of the frame starting from the initial position by a value of $\pi$ around the horizontal axis (the upside-down blue flower), and we call this horizontal reflection \textit{reflexión horizontal} $f_h$. Finally, the element $(c)$ is identified with the rotation of the frame starting from the initial position by a value of $\pi$ around the vertical axis (the blue flower). We call this vertical reflection \textit{reflexión vertical} $f_v$, as shown in Figure \ref{rotcuadroinic}.
     
    \begin{figure}[h]
        \centering
        \includegraphics[width=5cm, height=5cm]{fig/Ident-I.png}
        \hspace{0.5cm}
        \includegraphics[width=5cm, height=5cm]{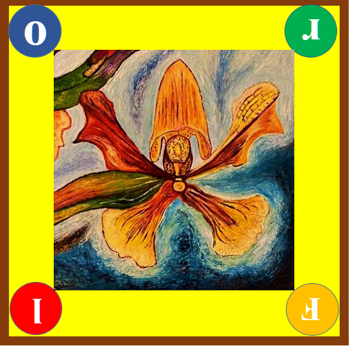} \\
        \vspace{0.5cm}
        \includegraphics[width=5cm, height=5cm]{fig/Ref-Fv.png}
        \hspace{0.5cm}
        \includegraphics[width=5cm, height=5cm]{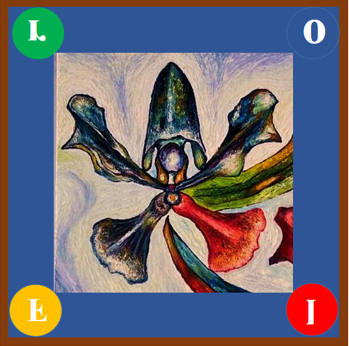}
        \caption{This figure shows the movements of the frame identified with the elements of the Klein group. The two upper frames correspond to rotation movements of the $(CI)$: the initial position $I$ (left) and the $\pi$ rotation, $R_2$ (right). At the bottom, the rotation movements of the $(CI)$ are the vertical reflection $f_v$ (left) and the horizontal reflection $f_h$ (right). Source: Author's elaboration.}
        \label{rotcuadroinic}        
    \end{figure}
        
    These identifications establish a correspondence between the rotation movements of the $(CI)$ and the operations of the Klein group. The rotation movements of the $(CI)$ satisfy the composition operations of the Klein group, expressed by the relations: 
   \begin{equation*}
        (R_2)^2=(f_v)^2=(f_h)^2=I, \;\;\;\;\; \;\;\;\;\; R_2f_h=f_v, \hspace{0.5cm}  f_h f_v = R_2.
\end{equation*}

The number $2$ in the exponent means to apply the corresponding rotation twice. For example, the operation  $ab=c$ in the Klein group equates to applying to the initial frame the rotation $R_2$ and then the horizontal reflection $f_h$, resulting in the vertical reflection $f_v$.
\end{enumerate}

\begin{flushleft}
\textbf{\textit{The Quaternion Group and the Initial Frame with Strings Tied  $(CIC)$}}
\end{flushleft}

Unlike the Klein group, to relate the quaternion group $\mathbb{Q}$ with the initial frame $(CI)$, it is necessary for the frame to have a belt or strings tied to one of its sides. Some authors refer to this as a puppet  \cite{diaz2017fenomeno, kauffman2001knots}. Incorporating strings into the initial frame allows for the exploration of new properties and behaviors of the system, generating more complex interactions and expanding the possibilities of movement and transformation of the frame.

\begin{figure}[h]
    \centering
    \includegraphics[width=5cm, height=7cm]{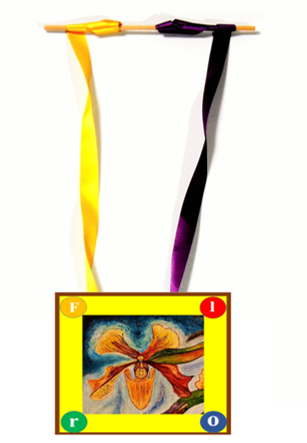}
    \caption{Frame in initial position with two tied strings hanging $(CIC)$, identified as $1\in \mathbb{Q}$. Source: Author's elaboration.}
    \label{cuadrocolg}        
\end{figure}

The identifications of the rotation movements of the $(CI)$ with tied strings establish a correspondence with the operations and cyclic permutations of the quaternion group $\mathbb{Q}$. To achieve this, we consider the $(CI)$ hung on the wall by two strings: a yellow string and a violet string, ensuring that the yellow flower is in the initial position (this initial position of the frame with strings tied is identified with the $1$ of $\mathbb{Q}$), as illustrated in Figure \ref{cuadrocolg}. This configuration of $(CI)$ with strings tied and hung $(CIC)$ provides a visual and tangible representation of the algebraic relationships present in the quaternion group.

\begin{figure}[h]
    \centering
    \includegraphics[width=6cm, height=5cm]{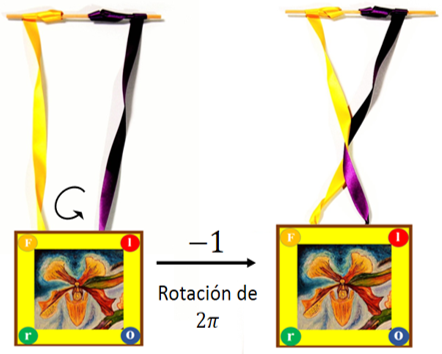}
    \hspace{1cm}
    \includegraphics[width=6cm, height=5cm]{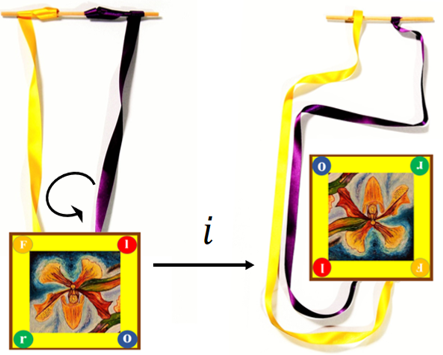}

    \vspace{1cm}
    \includegraphics[width=7cm, height=5cm]{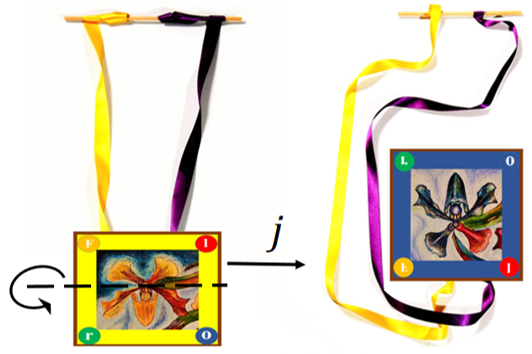}
    \hspace{1cm}
    \includegraphics[width=6cm, height=5cm]{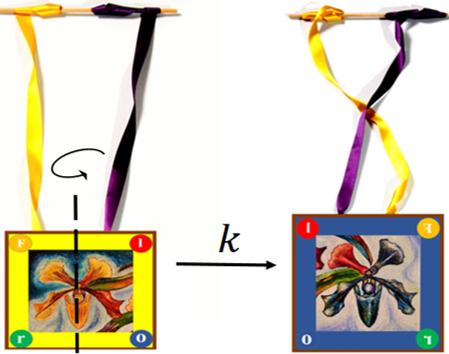}
    \caption{The quaternion group $\mathbb{Q}$ as rotation configurations, in a $\mathbb{R}^3$ frame with tied strings. Source: Author's elaboration.}
    \label{identicuadroinic}        
\end{figure}

The process of identification between the $(CIC)$ and the quaternion group $\mathbb{Q}$ is as follows: Starting with the $(CIC)$, we rotate the frame $2\pi$ with respect to the vertical axis, the frame returns to its original state, but the strings have a twist (they become entangled by one turn). We identify this position of the figure with the value of $-1$  (this convention is fundamental to achieving the relationship between the quaternion group $\mathbb{Q}$ and the rotation movements of the $(CIC)$). Again, starting with the $(CIC)$, we rotate a value of $\pi$ around the axis perpendicular to the plane of the frame that passes through the center of the frame, obtaining the upside-down yellow flower, and the strings do not have any twist. We identify this position of the frame with the value of $i$. Following the same process as before, we rotate a value of $\pi$ around the horizontal axis, obtaining the upside-down blue flower, and the strings cross. We identify this position of the frame with the value of $j$. Lastly, following the same process as the others, we rotate a value of $\pi$ around the vertical axis, obtaining the blue flower and the strings cross. We identify this position of the frame with the value of  $k$. Figure \ref{identicuadroinic} shows all the configurations described above.

\begin{enumerate}
    \item \textit{The Tangled $(CIC)$}
    
    An interesting phenomenon arises when considering the identifications of elements of the quaternion group $\mathbb{Q}$ with spatial rotations applied to the $(CIC)$. Applying the corresponding rotations $i^2, j^2$ y $k^2$ (where the superscript 2 indicates that the rotation is performed twice), the strings experience a twist of $2\pi$. This twist results in the strings being tangled, interpreted as a $2\pi$ twist and identified with $-1$. Consequently, we obtain the following equalities:
    $$i^2 = j^2 = k^2 = ijk = -1$$
    
    We now face the challenge of untangling the $(CIC)$ strings while maintaining consistency with the rotation movements and algebraic operations of the elements of the quaternion group $\mathbb{Q}$. To achieve this goal, we turn to the concept of the Dirac belt trick.
    
    The Dirac belt trick originates from the observation that when a spinor, such as an electron, completes a $2\pi$ rotation, its wave function in quantum mechanics inverts its sign, which has observable implications \cite{silverman1980curious}. However, a second $2\pi$ rotation restores the wave function to its original form \cite{staley2010understanding}. This phenomenon reveals that a $2\pi$ rotation is not equivalent to a null rotation but that a complete rotation requires $4\pi$. For a deeper understanding of this concept, consult \cite{bolker1973spinor, penrose1984spinors, hansen1994magic, stojanoska2008touching}.

    \item \textit{The Untangled $(CIC)$}
                
        Starting with the tangled $(CIC)$, after a $2\pi$ rotation about the vertical axis that leaves the strings tangled, applying the corresponding rotations again to the $(CIC)$, we obtain the following equalities:
        \begin{eqnarray*}
        &&(i^2)^2 = (j^2)^2 = (k^2)^2 = (-1)^2 = 1, \\
        &&((ik)^2)^2 = ((-j)^2)^2 = (j^2)^2 = 1, \\
        &&((kj)^2)^2 = ((-i)^2)^2 = (i^2)^2 = 1, \\
        &&((ji)^2)^2 = ((-k)^2 )^2 = (k^2)^2 = 1.
        \end{eqnarray*}
        In summary, the $(CIC)$ strings can be untangled, returning both the strings and the image to their initial position through a $4\pi$ rotation. The ability of the $(CIC)$ to return to its initial position after a $4\pi$ rotation is due to it meeting the criterion of the Dirac belt trick.

    To untangle the strings, the $(CIC)$ is kept fixed while the strings are moved, as shown in Figure \ref{desenredoCUIC}.
                
    \begin{figure}[h]
        \centering
        \includegraphics[width=3.5cm, height=7cm]{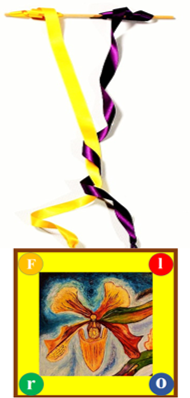}
        \hspace{0.5cm}
        \includegraphics[width=4cm, height=7cm]{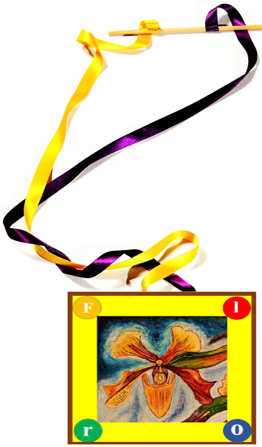}
        \hspace{0.5cm}
        \includegraphics[width=4cm, height=7cm]{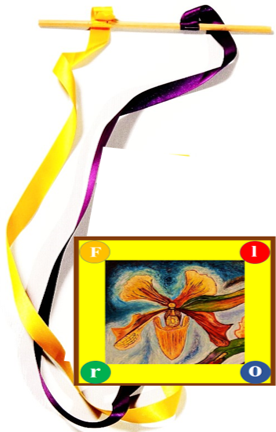} 
       \vspace{1cm}
        \includegraphics[width=4cm, height=7cm]{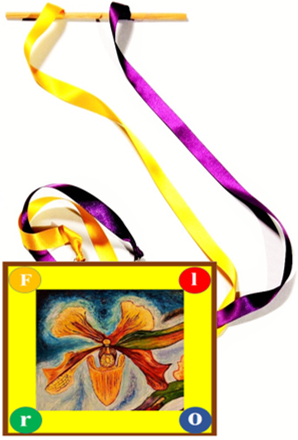}
        \hspace{0.5cm}
        \includegraphics[width=3.5cm, height=7cm]{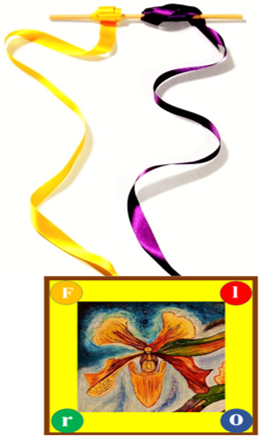}
        \hspace{1cm}
        \includegraphics[width=3.5cm, height=7cm]{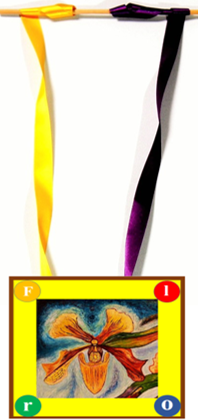} 
        \caption{Untangling the strings of the $CIC$ after a $4\pi$ rotation around the vertical axis.}
        \label{desenredoCUIC}        
    \end{figure}
\end{enumerate}

\textbf{Examples:} Multiplications in the Quaternion Group $\mathbb{Q}$ and rotations in the $(CIC)$.

\begin{enumerate}
\item The value $-i$ in the quaternion group $\mathbb{Q}$ is equivalent to performing a $2\pi$ twist in the strings and then applying the corresponding rotation to $i$, which is a $\pi$ rotation in the $(CIC)$ as shown in the left panel of Figure \ref{ejemplo-i}.

\begin{figure}[h]
            \centering
            \includegraphics[width=6cm, height=6cm]{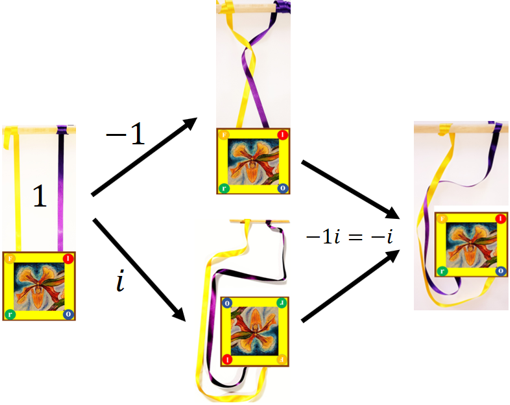} 
            \hspace{1cm}
            \includegraphics[width=6cm, height=8cm]{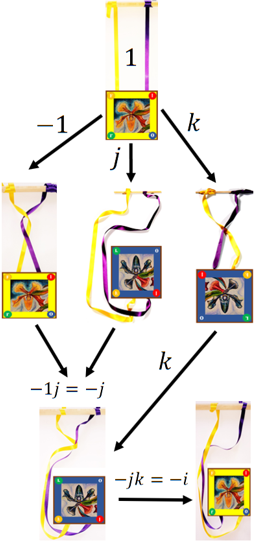}
            \caption{In the left panel, the rotations of the $CIC$ are shown, first $-1$ and then $i$, obtaining $-i$. In the right panel, the $(CIC)$ rotations are shown, first $-j$ and then $k$, obtaining $-jk=-i$.}
            \label{ejemplo-i}        
    \end{figure}

\item The operation $(-j)k=-jk=-i$ in the quaternion group  $\mathbb{Q}$ is equivalent to performing a $2\pi$ twist in the strings, followed by the application of the rotation assigned to $j$, which corresponds to a horizontal reflection in the $(CIC)$. Then, a vertical reflection is performed on the resulting position, causing a new twist in the strings and a $\pi$ rotation in the $(CIC)$, corresponding to $-i$. The visual representation of this process is shown in the right panel of Figure \ref{ejemplo-i}.
\end{enumerate}

\subsection{Quaternions and Pauli Matrices}

Pauli matrices, named after Wolfgang Ernst Pauli, are used in quantum physics to describe particles' intrinsic angular momentum or spin. Within the framework of quantum mechanics, these matrices appear in the Pauli equation, which describes the interaction of a particle's rotation with an external electromagnetic field. Each Pauli matrix is associated with an angular momentum operator, which can be classically interpreted as the rotations of elementary spin $\frac{1}{2}$ particles in each of the three spatial directions.

A connection between quaternions $\mathbb{H}$ and Pauli matrices is described as follows. 

To establish the relationship between quaternions $\mathbb{H}$ and Pauli matrices, consider the following $2\times 2$ matrix with complex number entries: 
\begin{equation*}
       L=
        \begin{bmatrix}
        a+d & b-ic\\
        b+ic & a-d\\
        \end{bmatrix}
\end{equation*}
con $a, b, c \in \mathbb{R}$ and $i^2=-1$. Observe that the conjugate matrix

\begin{equation*}
       L^{\ast}=
        \begin{bmatrix}
        a+d & b+ic\\
        b-ic & a-d\\
        \end{bmatrix}
\end{equation*}
and the conjugate transpose matrix $(L^{\ast})^T$:
\begin{equation*}
       (L^{\ast})^T=
        \begin{bmatrix}
        a+d & b-ic\\
        b+ic & a-d\\
        \end{bmatrix}
        =L
\end{equation*}
fulfill the condition $L=(L^{\ast})^T$; it also holds that $L=(L^T)^{\ast}$, implying that $L$ is an Hermitian matrix.

Moreover, the determinant of matrix $L$ is defined as: 
\begin{equation*} 
    \text{det}(L) = (a+d)(a-d) - (b+ic)(b-ic) = (a^2 - d^2) - (b^2 + c^2) = a^2 - b^2 - c^2 - d^2. 
\end{equation*}

The set of all matrices with the characteristics of $L$ is known as the set of Lorentz matrices $ML$. In physics, quaternions are related to the nature of the universe at the level of quantum mechanics \cite{jia2008quaternions}, providing elegant expressions for Lorentz transformations, fundamental in the modern theory of relativity \cite{freund2008special}

We define the function $f: \mathbb{R}^4 \longrightarrow ML$ with the rule of correspondence:
\begin{equation*}
   f(v) = f(a,b,c,d) =
    \begin{bmatrix}
    a+d & b-ic\\
    b+ic & a-d\\
    \end{bmatrix}
\end{equation*}
where $v = (a,b,c,d) \in \mathbb{R}^4$. This function fulfills the condition of being linear:
\begin{equation*}
    f(rv+sw) = rf(v) + sf(w), 
\end{equation*}
for $r, s \in \mathbb{R}$ and $v, w \in \mathbb{R}^4$.

Applying the function $f$ to the canonical basis $\{e_0=(1,0,0,0), e_1=(0,1,0,0), e_2=(0,0,1,0), e_3=(0,0,0,1)\}$ of the space $\mathbb{R}^4$, the following matrices are obtained:
\begin{equation*}
    f(e_0) = \begin{bmatrix} 1 & 0 \\ 0 & 1 \end{bmatrix} = \sigma_0, \quad
    f(e_1) = \begin{bmatrix} 0 & 1 \\ 1 & 0 \end{bmatrix} = \sigma_1 = \sigma_x
\end{equation*}

\begin{equation*}
    f(e_2) = \begin{bmatrix} 0 & -i \\ i & 0 \end{bmatrix} = \sigma_2 = \sigma_y, \quad
    f(e_3) = \begin{bmatrix} 1 & 0 \\ 0 & -1 \end{bmatrix} = \sigma_3 = \sigma_z
\end{equation*}

The Hermitian matrices $\sigma_1, \sigma_2$ and $\sigma_3$ are known as the Pauli matrices, while $\sigma_0$ is simply the identity matrix. 

Multiplying the Pauli matrices by the complex number $i$, the space generated by the set  $P = \{\sigma_0, i\sigma_1, i\sigma_2, i\sigma_3\}$ turns out to be isomorphic to the quaternions $\mathbb{H}$. The proof of this assertion is outlined below.

We define the function $h: \mathbb{H} \longrightarrow P$ with the following rule of correspondence:
\begin{eqnarray*}
    h(q) &=& h(a+bi+cj+dk) = a\sigma_0 - bi\sigma_1 - ci\sigma_2 - di\sigma_3, \\
    h(q) &=& a\sigma_0 - i(b\sigma_1 + c\sigma_2 + d\sigma_3),
\end{eqnarray*}
where $q = a+ib+cjdk \in \mathbb{H}$. The function $h$ meets the following conditions:
\begin{enumerate}

    \item 
    \begin{equation*}
     h(a+bi+cj+dk) = a
    \begin{bmatrix}
    1 & 0\\
    0 & 1\\
    \end{bmatrix}
   - i(b
    \begin{bmatrix}
    0 & 1\\
    1 & 0\\
    \end{bmatrix}
    + c
    \begin{bmatrix}
    0 & -i\\
    i & 0\\
    \end{bmatrix}
    + d
    \begin{bmatrix}
    1 & 0\\
    0 & -1\\
    \end{bmatrix}
    )
   \end{equation*}

   \begin{equation*}
     h(a+bi+cj+dk) = 
    \begin{bmatrix}
    a & 0\\
    0 & a\\
    \end{bmatrix}
   + 
    \begin{bmatrix}
    0 & -ib\\
    -ib & 0\\
    \end{bmatrix}
    +
    \begin{bmatrix}
    0 & -c\\
    c & 0\\
    \end{bmatrix}
    +
    \begin{bmatrix}
    -id & 0\\
    0 & id\\
    \end{bmatrix}
    \end{equation*}

    \begin{equation*}
     h(a+bi+cj+dk) = 
    \begin{bmatrix}
    a-id & -(c+ib)\\
    c-ib & a+id\\
    \end{bmatrix}
   \end{equation*}
    
    \item The function $h$ is linear, i.e., $h(rq_1+sq_2)=rh(q_1)+sh(q_2)$, for $r,s\in \mathbb{R}$ and $q_1,q_2 \in\mathbb{H}$.
    
    \item The function $h$ applied to the quaternions $1,i,j,k \in \mathbb{H}$ satisfies the following equalities.

    \begin{equation*}
     h(1) = 1\sigma_0 - i(0\sigma_1 + 0\sigma_2 + 0\sigma_3) =
    \begin{bmatrix}
     1 & 0\\
     0 & 1\\
    \end{bmatrix}
    = \sigma_0
    \end{equation*}
    
    \begin{equation*}
     h(i) = 0\sigma_0 - i(1\sigma_1 + 0\sigma_2 + 0\sigma_3) = -i
    \begin{bmatrix}
     0 & 1\\
     1 & 0\\
    \end{bmatrix}
    = -i\sigma_1
   \end{equation*}

    \begin{equation*}
     h(j) = 0\sigma_0 - i(0\sigma_1 + 1\sigma_2 + 0\sigma_3) = -i
    \begin{bmatrix}
     0 & -i\\
     i & 0\\
    \end{bmatrix}
    = -i\sigma_2
   \end{equation*}

     \begin{equation*}
     h(k) = 0\sigma_0 - i(0\sigma_1 + 0\sigma_2 + 1\sigma_3) = -i
    \begin{bmatrix}
     1 & 0\\
     0 & -1\\
    \end{bmatrix}
    = -i\sigma_3
   \end{equation*}
        
    \item The properties of the equality $i^2=j^2=k^2=ijk=-1$ in quaternions, the equivalents in the space $P$ are met.
    $$ (-i\sigma_1)^2 = (-i\sigma_2)^2 = (-i\sigma_3)^2 = -\sigma_0 $$
    
    \item The function $h$ is bijective.

\end{enumerate}

\textbf{Application of Pauli Matrices.} 

In the application of quaternions and rotations, the unit quaternion group $S^3$ describes three-dimensional rotations in $3D$ space. There is a way to relate this group to the Pauli matrices.

Define the function $g: S^3 \longrightarrow M_{2\times2}(\mathbb{C})$ by the rule of correspondence that is a linear combination of the Pauli matrices multiplied by the complex $i$:

\begin{equation*}
g(q) = g(a+bi+cj+dk) = d\sigma_0 + ai\sigma_1 + bi\sigma_2 + ci\sigma_3
\end{equation*}

\begin{equation*}
g(a+bi+cj+dk) = d
\begin{bmatrix}
1 & 0\\
0 & 1\\
\end{bmatrix}
+ ia
\begin{bmatrix}
0 & 1\\
1 & 0\\
\end{bmatrix}
+ ib
\begin{bmatrix}
0 & -i\\
i & 0\\
\end{bmatrix}
+ ic
\begin{bmatrix}
1 & 0\\
0 & -1\\
\end{bmatrix}
\end{equation*}

\begin{equation*}
g(a+bi+cj+dk) = 
\begin{bmatrix}
d & 0\\
0 & d\\
\end{bmatrix}
+ 
\begin{bmatrix}
0 & ia\\
ia & 0\\
\end{bmatrix}
+
\begin{bmatrix}
0 & b\\
-b & 0\\
\end{bmatrix}
+
\begin{bmatrix}
ic & 0\\
0 & -ic\\
\end{bmatrix}
\end{equation*}

\begin{equation*}
g(a+bi+cj+dk) = 
\begin{bmatrix}
d+ic & b+ia\\
-b+ia & d-ic\\
\end{bmatrix}
= C
\end{equation*}

The matrix $C$ satisfies the following conditions:

\begin{enumerate}
    \item The determinant of the matrix $C$ is equal to 1:
    \begin{align*}
    \text{det}(C) &= (d+ic)(d-ic)-(-b+ia)(b+ia) \\
    &= d^2+c^2-(-b^2-a^2) \\
    &= a^2+b^2+c^2+d^2 = 1
    \end{align*}
    
    \item The conjugate transpose matrix $(C^{\ast})^T$ is the inverse of $C$, i.e., the equality $C(C^{\ast})^T=I$ is met:
    \begin{align*}
    C^{\ast} &= 
    \begin{bmatrix}
    (d+ic)^{\ast} & (b+ia)^{\ast}\\
    (-b+ia)^{\ast} & (d-ic)^{\ast}\\
    \end{bmatrix}
    = \begin{bmatrix}
    d-ic & b-ia\\
    -b-ia & d+ic\\
    \end{bmatrix}
    \end{align*}
    
    \begin{align*}
    (C^{\ast})^T &= 
    \begin{bmatrix}
    d-ic & -b-ia\\
    b-ia & d+ic\\
    \end{bmatrix}
    \end{align*}
    
    \begin{align*}
    C(C^{\ast})^T &= 
    \begin{bmatrix}
    d+ic & b+ia\\
    -b+ia & d-ic\\
    \end{bmatrix}
    \begin{bmatrix}
    d-ic & -b-ia\\
    b-ia & d+ic\\
    \end{bmatrix} \\
    &= \begin{bmatrix}
    d^2+c^2+b^2+a^2 & 0\\
    0 & b^2+a^2+d^2+c^2\\
    \end{bmatrix} = I
    \end{align*}
    
In other words, the matrix $C$ is unitary.
\end{enumerate}

The set of all matrices $C$ with the above characteristics, with entries in complex numbers $(\mathbb{C})$, unitary and with determinant 1, is called the special unitary matrix group or special unitary group $SU(2)$. The group operation is given by matrix multiplication. It's a group because, for two matrices $C, B \in SU(2)$, the product is again a matrix in $SU(2)$. Furthermore, the product of matrices is associative, every matrix $C$ has an inverse matrix $(C^{\ast})^T$, and lastly the identity matrix is:

\begin{equation*}
I=
\begin{bmatrix}
1 & 0\\
0 & 1\\
\end{bmatrix}
\end{equation*}

Restricting the codomain of the function $g$ to the group  $SU(2)$, $g: S^3 \longrightarrow SU(2)$ with the same rule of correspondence:

\begin{equation*}
g(q) = g(a+bi+cj+dk) = 
\begin{bmatrix}
d+ic & b+ia\\
-b+ia & d-ic\\
\end{bmatrix}
= C
\end{equation*}

$g$ is an isomorphism. In other words, the group $S^3$ is isomorphic to the special unitary group $SU(2)$.

So far, we have developed the interrelationship between quaternions and Pauli matrices, revealing a deep and fundamental connection between these two mathematical tools in contemporary theoretical physics. Through the defined function $g$, we have established an isomorphism between the group of unitary quaternions $S^3$ and the unitary special group $SU(2)$ of Pauli matrices. This algebraic connection provides a deeper understanding of three-dimensional rotations, demonstrating how the abstract principles of linear algebra and group theory underlie concrete physical phenomena. This understanding reinforces our perception of how fundamental mathematical concepts find practical applications in the description and understanding of nature.

\section{Conclusions}

Throughout this article, we have explored the fascinating history of quaternions, from their invention by Sir William Rowan Hamilton to their impact on various scientific and technological disciplines. 

Our work presents significant advancements compared to \cite{mcdonald2010teaching}. In the context of rotation applications, we thoroughly address quaternions, showcasing their representation in complex numbers and as matrices and distinguishing between right and left rotations through detailed numerical examples. In contrast, previous works focus only on the second matrix form of quaternions without exploring the diversity of representations and applications. Moreover, we propose a general approach to rotations through a theorem and a clear definition of the associated morphism, providing a solid foundation for understanding the algebraic structure of quaternions. We also highlight the distinction between left and right rotations and discuss their physical significance, aspects generally not addressed in previous works.

An innovative aspect of our work is presenting a three-dimensional example that illustrates the rotation of a picture with strings, linking quaternions with the quaternion group, the phenomenon of half-integer spin, and the Pauli matrices. This approach offers a unique perspective that connects advanced theoretical concepts with practical applications, significantly contributing to understanding and applying quaternions in various scientific and technological fields.
Finally, we emphasize the importance of scientific outreach in making complex concepts such as quaternions accessible. Public understanding of these topics contributes to advancing knowledge and stimulates future research.

\section*{Acknowledgements}

We would like to acknowledge the funding provided by the projects SIP20230505 and SIP20240638 of the IPN Secretaría de Investigación y Posgrado, as well as the FORDECYT-PRONACES-CONACYT CF-MG-2558591 project, which contributed to the development of this work. FRGD sincerely acknowledges Reynaldo González Díaz and José Macario Moreno Calzada for their valuable support.

\bibliographystyle{unsrt}
\bibliography{Cuaterniones}

\end{document}